\documentclass{report}

\usepackage{amsmath}
\usepackage{amssymb}
\usepackage{graphicx}
\usepackage{lipsum}
\usepackage{hyperref}
\usepackage{mathtools, nccmath}
\usepackage{longtable}
\usepackage[numbers,square]{natbib}
\usepackage{aastex_macros}
\usepackage{authblk}

\title{
\begin{figure}
    \centering
    \includegraphics[width=1.0\linewidth]{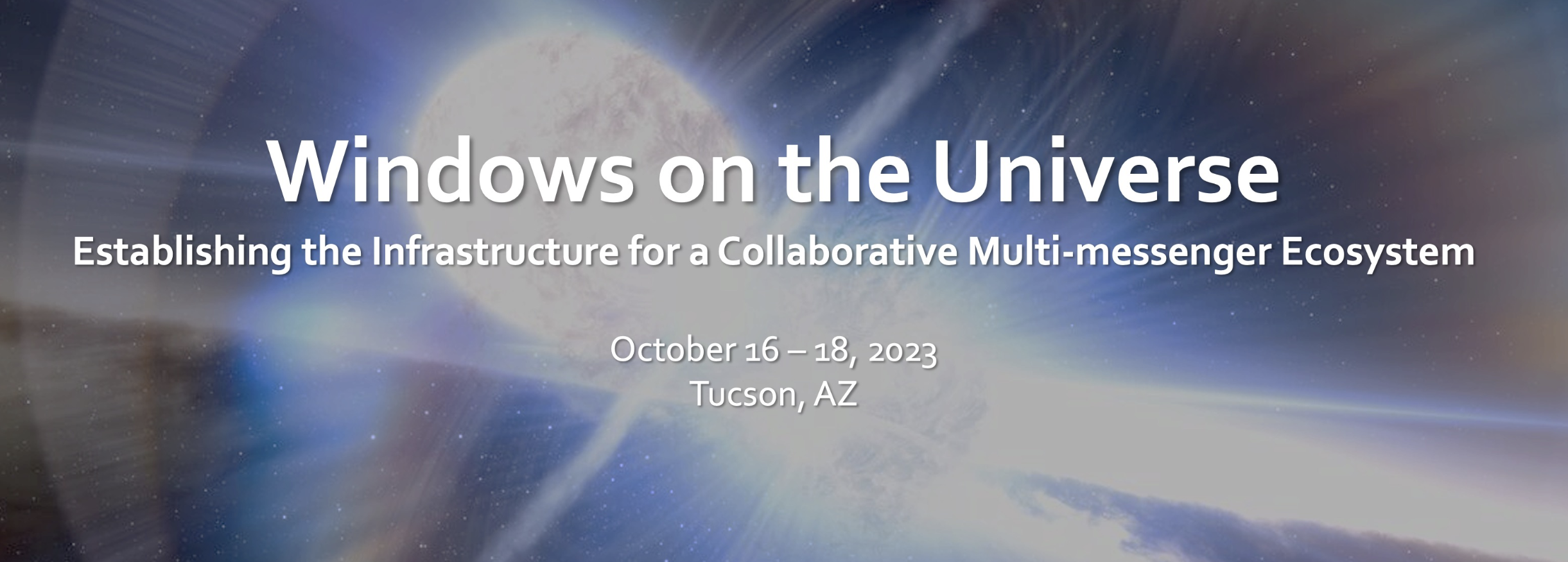}
\end{figure}
\clearpage
}

\author[1]{The 2023 Windows on the Universe Workshop White Paper Working Group: Tom\'as Ahumada}
\author[2,3]{Jennifer E. Andrews (SOC co-chair)}
\author[4]{Sarah~Antier}
\author[5]{Erik Blaufuss}
\author[2,6]{P.R.~Brady}
\author[7]{A.M.~Brazier}
\author[8]{Eric Burns}
\author[2,9,10]{S.~Bradley Cenko (editor)}
\author[11]{Poonam Chandra} 
\author[12]{Deep Chatterjee}
\author[2,13]{Alessandra Corsi}
\author[14]{Michael W. Coughlin}
\author[15]{David~A.~Coulter}
\author[16]{Shenming Fu}
\author[2,17]{Adam~Goldstein}
\author[16]{Leanne P. Guy}
\author[18]{Eric J. Hooper}
\author[19]{Steve~B.~Howell}
\author[9]{T.B.~Humensky}
\author[20]{Jamie A. Kennea}
\author[21]{S.M.~Jarrett}
\author[2,16]{Ryan M.~Lau (SOC co-chair)}
\author[22]{Tiffany R. Lewis}
\author[23]{Lu Lu}
\author[16]{Thomas Matheson}
\author[24]{Bryan W. Miller}
\author[25]{Gautham Narayan}
\author[16]{Robert Nikutta}
\author[16]{Jayadev K. Rajagopal}
\author[2,15]{Armin Rest}
\author[26]{K.M.~Ruiz-Rocha}
\author[2,26]{Jessie Runnoe} 
\author[2,27]{David J. Sand (editor)}
\author[28]{Marcos Santander}
\author[2,20]{Hugo~A.~Ayala~Solares}
\author[2,3]{Monika~D.~Soraisam (SOC co-chair)}
\author[2,29]{R.A.~Street}
\author[30]{Aaron Tohuvavohu}
\author[25]{Joaquin~Vieira}
\author[34]{Abigail~Vieregg}
\author[6]{Sarah~J.~Vigeland}
\author[31]{Salvatore Vitale}
\author[32]{Nicholas~E.~White}
\author[33]{Samuel D. Wyatt}
\author[23]{Tianlu Yuan}

\affil[1]{Division of Physics, Mathematics and Astronomy, California Institute of Technology, Pasadena, CA 91125, USA} 
\affil[2]{Scientific Organizing Committee} 
\affil[3]{Gemini Observatory/NSF's NOIRLab, 670 N. A'ohoku Place, Hilo, HI 96720, USA} 
\affil[4]{Observatoire de la C\^ote d'Azur, Universit\'e C\^ote d'Azur, Boulevard de l'Observatoire, 06304 Nice, France} 
\affil[5]{Department of Physics, University of Maryland, College Park, MD 20742, USA} 
\affil[6]{Center for Gravitation, Cosmology and Astrophysics, University of Wisconsin--Milwaukee, P.O. Box 413, Milwaukee WI, 53201, USA} 
\affil[7]{Cornell University, Ithaca, NY 14853} 
\affil[8]{Department of Physics \& Astronomy, Louisiana State University, Baton Rouge, LA 70803, USA}
\affil[9]{Astrophysics Science Division, NASA Goddard Space Flight Center, Greenbelt, MD 20771, USA} 
\affil[10]{Joint Space-Science Institute, University of Maryland, College Park, MD 20742 USA} 
\affil[11]{National Radio Astronomy Observatory, 520 Edgemont Rd, Charlottesville VA 22903, USA} 
\affil[12]{LIGO Laboratory and Kavli Institute for Astrophysics and Space Research, Massachusetts Institute of Technology, 185 Albany Street, Cambridge, Massachusetts 02139, USA}
\affil[13]{Department of Physics and Astronomy, Texas Tech University, Box 1051, Lubbock, TX 79409-1051, USA}
\affil[14]{School of Physics and Astronomy, University of Minnesota, Minneapolis, Minnesota 55455, USA} 
\affil[15]{Space Telescope Science Institute, 3700 San Martin Drive, Baltimore, MD 21218-2410, USA} 
\affil[16]{NSF’s National Optical-Infrared Astronomy Research Laboratory, 950 North Cherry Avenue, Tucson, AZ 85719, USA} 
\affil[17]{Science and Technology Institute, Universities Space Research Association, Huntsville, AL 35805, USA} 
\affil[18]{Astronomy Department, University of Wisconsin-Madison, 475 N. Charter St., Madison, WI 53706, USA} 
\affil[19]{NASA Ames Research Center, Moffett Field, CA 94035, USA}
\affil[20]{Department of Astronomy and Astrophysics, The Pennsylvania State University, 525 Davey Lab, University Park, PA 16802, USA} 
\affil[21]{Vanderbilt University, 2201 West End Ave, Nashville, TN 37209, USA} 
\affil[22]{Department of Physics, Earth, Planetary \& Space Sciences Institute, Michigan Technological University, Houghton, MI 49931, USA} 
\affil[23]{Department of Physics \& Wisconsin IceCube Particle Astrophysics Center,
University of Wisconsin–Madison, Madison, WI 53706, USA} 
\affil[24]{Gemini Observatory/NSF's NOIRLab, Casilla 603, La Serena, Chile}
\affil[25]{Department of Astronomy, University of Illinois, Urbana-Champaign, 1002 W. Green St., Urbana, IL 61801, USA} 
\affil[26]{Department of Physics and Astronomy, Vanderbilt University,2201 West End Ave, Nashville, TN 14853}
\affil[27]{Steward Observatory, University of Arizona, 933 North Cherry Avenue, Tucson, AZ 85721-0065, USA} 
\affil[28]{Department of Physics and Astronomy, University of Alabama, Tuscaloosa, AL 35487, USA} 
\affil[29]{Las Cumbres Observatory, 6740 Cortona Drive, Suite 102, Goleta, CA 93117, USA} 
\affil[30]{Department of Astronomy \& Astrophysics, University of Toronto, Toronto, ON M5S 3H4} 
\affil[31]{Department of Physics, Massachusetts Institute of Technology, 77 Massachusetts Avenue, Cambridge, MA 02139, USA} 
\affil[32]{Department of Physics, The George Washington University, Corcoran Hall, 725 21st Street NW, Washington DC 20052, USA} 
\affil[33]{Department of Astronomy, University of Washington, 3910 15th Avenue NE, Seattle, WA 98195-0002, USA} 
\affil[34]{University of Chicago, Chicago, IL 60637, USA} 

\begin{document}

\maketitle

\tableofcontents

\chapter{Executive Summary}
\label{sec:ES}
The field of multi-messenger and time-domain astronomy (MMA/TDA) is poised for breakthrough discoveries in the coming decade. In much the same way that expanding beyond the optical bandpass revealed entirely new and unexpected discoveries, cosmic messengers beyond light (i.e., gravitational waves [GWs], neutrinos, and cosmic rays) open entirely new windows to answer some of the most fundamental questions in (astro)physics: Where are heavy elements synthesized? What is the equation of state of dense matter? What is the cosmology of our Universe? What caused the generation of matter over antimatter? How does particle acceleration work? Furthermore, observational capabilities in this field are advancing at an almost unprecedented pace: for example, it is conceivable with the Cosmic Explorer experiment\footnote{\url{https://cosmicexplorer.org}} that the field of high-frequency GWs will go from the discovery of the first stellar-mass binary black hole merger in 2015 \citep{GW150914} to sufficient sensitivity to detect \textbf{all} such systems in the Universe in just a few decades. For all of these reasons, MMA/TDA was prioritized as a frontier scientific pursuit in the 2020 Decadal Survey on Astronomy and Astrophysics via its ``New Windows on the Dynamic Universe'' theme.

At the same time, MMA/TDA science presents technical challenges distinct from those experienced in other disciplines. Successful observations require coordination across myriad boundaries -- different cosmic messengers, ground vs.~space, international borders, etc. -- all for sources whose sky position may not be well localized, and whose brightness may be changing rapidly with time. Theoretical predictions of anticipated behavior require consideration of a broad range of fundamental theories (general relativity, magnetic fields, complex radiation transport) across a large dynamic range of spatial and temporal scales. And all of this work is undertaken by real human beings, with distinct backgrounds, experiences, cultures, and expectations, that often conflict.

To address these challenges and help MMA/TDA realize its full scientific potential in the coming decade (and beyond), the second in a series of community workshops sponsored by the U.S. National Science Foundation (NSF) and NASA titled ``Windows on the Universe: Establishing the Infrastructure for a Collaborative Multi-Messenger Ecosystem'' was held on October 16-18, 2023 in Tucson, AZ\footnote{Because of the workshop organization/structure, we note that most of the recommendations here are U.S.-centric. We do attempt to expand beyond this remit at times, and hope that international coordination will be a focus of future MMA/TDA workshops.}. Here we summarize the primary recommendations from this workshop. Four key questions to address were identified prior to the workshop:
\begin{itemize}
    \item What are the main challenges to perform successful MMA/TDA campaigns and to maximize their scientific potential?
    \item How should we coordinate MMA/TDA follow-up to reduce operational redundancy across the network of ground and space-based observatories?
    \item How should we foster collaboration in the MMA/TDA community?
    \item How can we ensure that the MMA/TDA field reaches its full potential over the next decade?
\end{itemize}

\section{Hardware Recommendations}
MMA/TDA is inherently a discovery-based science. While theory is critical both to guide observers in which direction(s) to look and to fully maximize scientific inference from undertaken campaigns, \textbf{a robust suite of observational facilities underlies all of MMA/TDA science.}

There is no single instrument or facility that can address the breadth of MMA/TDA science. Rather, a vibrant \textbf{ecosystem} is required, with engines for discovery and follow-up/characterization required across the electromagnetic spectrum and beyond. Furthermore, in some areas upcoming/upgraded facilities promise revolutionary breakthroughs (e.g., Rubin, Roman, LIGO A+, LISA, IceCube-Gen2, Cosmic Explorer), while in others future prospects are less clear (e.g.,  aging of Swift and Fermi, ground-based optical spectroscopy). Considering broadly this landscape, we make the following recommendations:
\begin{enumerate}
    \item \textbf{To NASA:} With the advanced age of many of the key MMA/TDA assets in its fleet, we urge NASA to rapidly develop an implementation plan for the highest priority sustaining capability for space identified in the 2020 Decadal Survey:

    \textit{``To advance [MMA/TDA] science, it is essential to maintain and expand space-based time-domain and follow up facilities ... Many of the necessary observational capabilities can be realized on Explorer-scale platforms, or possibly somewhat larger. As the international landscape and health of NASA assets change, it will be important for NASA to seek regular advice over the coming decade on needed capabilities and to ensure their development. The open Explorer program calls have reached a healthy funding level, and ... maintaining the current cadence of open calls is a condition for new initiatives. This time-domain program is therefore recommended as an augmentation to those levels, and would be executed through competed calls in broad, identified areas.''}

    We recognize the programmatic and fiscal challenges such a program entails, but an implementation plan -- including a prioritized list of needed capabilities and a timeline for competed MMA/TDA mission calls -- is both readily achievable and extremely urgent. The proposal process for future space missions must establish a means to enable coordinated campaigns with non-NASA facilities in order to achieve key Decadal MMA/TDA priorities. Wide-field, high-energy monitoring and rapid-response UV and X-ray follow-up are clearly the highest priorities for the community, and facilities must be designed strategically from the outset in order to work effectively in concert (e.g., wide-field monitor localization capability matched to follow-up facility field-of-view).
    \item \textbf{To NSF Astronomy:} At optical/NIR wavelengths, continued investment in wide-field imaging is critical for a broad range of MMA/TDA science -- for example, the optical/NIR kilonova signal appears to be the most plausible EM counterpart of a binary neutron star merger for most viewing angles. \textbf{Rubin will be revolutionary in this regard, but it is not sufficient -- higher cadence, broader wavelength and full sky coverage (e.g., Northern and Southern hemispheres) cannot be achieved with a single telescope.} Sustained support for new and/or existing wide-area imaging facilities will be necessary even in the Rubin era.

    Follow-up spectroscopy -- already a bottleneck for the community -- will only become more of a problem in the next decade. A significant investment in new \textbf{community accessible} optical/NIR spectroscopy (e.g., X-shooter \cite{Xshooter}; SCORPIO \cite{SCORPIO}) is desperately needed. Absent this support, a significant fraction of MMA/TDA science possible with bedrock facilities such as Rubin and LIGO will go unrealized.   

    At radio wavelengths, the highest priority is continued support for the next-generation VLA (ngVLA\footnote{\url{https://ngvla.nrao.edu/}}), which provides the sensitivity and resolution necessary to match corresponding gains made by, e.g., ground-based GW detectors.

    NSF should develop, in coordination with NASA, a policy that supports future MMA/TDA space mission proposal observing time requirements.
    
    \item \textbf{To NSF Physics:} We strongly endorse advisory recommendations from the 2020 Decadal Survey, including support for robust ``investments in technology development for advanced 
    gravitational wave interferometers, both to upgrade LIGO and to prepare for the next large facility,'' as well as for the development of IceCube-Gen2. Continued sensitivity improvements in ground-based GW and high-energy neutrino detectors are vital given the astrophysical rates / density of relevant MMA sources; however, it is critical these sensitivity upgrades be achieved on a reliable schedule, avoiding extended down periods as much as possible, in order to maximize joint observing campaigns and maintain a robust follow-up community.     
\end{enumerate}

\section{Software Recommendations}
The unique challenges of MMA/TDA -- coordinating observations across multiple cosmic messengers in real time -- require diverse software tools and services. While the community has made great gains in the software area in recent years (e.g., SCiMMA, GCN, Treasure Map, TOM Toolkit), \textbf{software development remains under-resourced by funding agencies and under-valued by our community as a whole.} Thus, our first and foremost recommendation is a \textbf{significant increase in funding opportunities for MMA/TDA software} -- the current level of support is holding back scientific productivity in a number of areas.

In addition to inadequate overall support, the structure of most funding calls does not align well with the typical MMA/TDA development process. Seed funding for new ideas is largely absent, and \textbf{sustained support is extremely rare}. We strongly recommend a \textbf{phased} approach along the following lines:
\begin{itemize}
    \item \textit{Phase 1 (Alpha release for an individual/team):} many MMA/TDA software projects start with an early career scientist(s) identifying a concrete need on their research team/project, and developing a tool to address this. To meet this need, we recommend a continuously open solicitation for early career (graduate students and postdocs) researchers to fund 6-12 months of effort ($\sim \$ 100$\,k) on MMA/TDA-specific cyberinfrastructure. To ensure a diverse ecosystem of software tools, the success rate of these proposals should be purposefully high ($> 50$\%).
    \item \textit{Phase 2 (Beta release for the broader community):} The next step in the development cycle is to deploy successful tools beyond an individual research group/team to the broader community. For projects that meet their Phase 1 milestones, a cadenced (annual/bi-annual) call for Phase 2 proposals would provide sufficient resources for professional software development and maintenance ($\sim \$1$M). Sustained support (3--5 years) is critical to the success of such efforts. 
    \item \textit{Phase 3 (Production release for critical infrastructure):} Some software tools are so critical to the community that they cannot be allowed to be subject to the whims of annual peer review for funding (e.g., GCN, LSST-scale brokers, LVK alerts, science platforms). Funding agencies should define a process to identify the most vital MMA/TDA cyberinfrastructure, as well as possible paths for long-term (decade+) stability for these products. One potential solution is for such efforts to be redirected to a national lab, NASA center, or equivalent (e.g., NOIRLab), though other possibilities may also exist. 
    \end{itemize}

Finally, we reiterate that the interoperability of these tools and services, including interoperability between cross-agency and international archives, is essential for a robust MMA/TDA software ecosystem. In some cases, joint NASA+NSF software development may reduce duplication of effort \textbf{and} improve community usability (e.g., joint high-energy, GW, and neutrino signal-driven alert searches) -- we encourage funding agencies to creatively pursue such opportunities. Furthermore, a dedicated MMA/TDA archive infrastructure that can bring together data from space- and ground-based observatories is also highly desirable.

\section{People and Policy Recommendations}
People and policy are important cornerstones for the MMA/TDA pursuit. To ensure the MMA/TDA field reaches its full potential, we make the following recommendations:

\begin{itemize}
\item Given the tremendous importance of infrastructure for the MMA/TDA enterprise, the entire community must provide improved career opportunities for those responsible for developing such software. Funding channels with industry-competitive salaries must be provided to enable software engineers to be employed in research groups long term ($5+$ years). Rewarding career paths must be enabled to retain software engineers in research as well as researchers with an interest in software development. Funding agencies should signal greater respect for researchers with this specialization in all agency presentations to the community, and senior researchers must weight software contributions as well as publications in tenure evaluation, fellowship selection panels, etc.

\item Closer coordination between NASA and the NSF is a \textbf{necessity}, particularly in long-term planning (e.g., to ensure contemporaneous operations of flagship facilities) and in negotiations with international partners.  The agencies should support existing efforts to enable MMA/TDA science at NASA and NSF observing facilities (e.g., ACROSS and AEON Network), and fund the extension of these initiatives to radio and other wavelength facilities.  

\item Expand community access to MMA/TDA data sets. For the most exceptional sources (e.g., next nearby GW counterpart, Galactic supernova), the MMA/TDA community should work with observatory staff to predefine observing programs whose data would become publicly available immediately upon triggering. For other sources, proprietary periods should be made as short as possible to maximize the scientific yield from these observations.
\end{itemize}

\chapter{Introduction}
\label{sec:intro}
Multi-messenger and time-domain astronomy (MMA/TDA) was prioritized as one of three frontier discovery areas in the 2020 Decadal Survey on Astronomy and Astrophysics. In large part the promise of this field results from the advent of powerful new (and upgraded) facilities probing cosmic messengers beyond light. Prominent examples include:
\begin{itemize}
    \item \textit{Ground-based Gravitational Waves (GWs):} The International GW Observatory Network [IGWN], comprised of the LIGO, Virgo, and KAGRA GW interferometers, directly detected GWs from a binary black hole merger for the first time in 2015 \citep{GW150914}. This discovery was sufficiently transformative for three LIGO founders to be awarded the 2017 Nobel Prize in Physics. Planned upgrades to the current network of detectors will result in nearly an order of magnitude increase in detection rate by the end of this decade, with even more powerful new facilities under consideration for the 2030's.
    \item \textit{High-Energy Neutrinos:} The IceCube Neutrino Observatory discovered a diffuse background of high-energy astrophysical neutrinos in 2013 \citep{HEN2013}. Subsequent data collection has revealed the active galaxy NGC\,1068 as likely the first (extrasolar) source of high-energy neutrinos just last year \citep{NGC1068}. Planned upgrades would increase the neutrino detection rate by an order of magnitude in the next decade \citep{IceCube-Gen2}.
    \item \textit{Pulsar Timing Arrays (PTAs):} PTA experiments around the world recently reported the first evidence for a gravitational wave background at nanohertz frequencies \citep{2023ApJ...951L...8A,2023A&A...678A..50E,2023ApJ...951L...6R,2023RAA....23g5024X}. The observed signal is consistent with a gravitational wave background produced by a population of supermassive binary black holes, although more exotic sources cannot be ruled out \citep{2023ApJ...951L..11A,2023ApJ...952L..37A,2023arXiv230616227A,2023PhRvL.131q1001S}.  In the future, PTAs are expected to observe GWs from individual supermassive binary black holes \citep{2017NatAs...1..886M,2018MNRAS.477..964K,2022ApJ...941..119B}. 
    \item \textit{Space-Based Gravitational-Wave Detection:} The upcoming Laser Interferometer Space Antenna (LISA) mission will be able to detect GWs between the high frequency ranges of ground-based GW detectors and the low frequencies probed by PTAs. Within this intermediate range should lie GW signatures from merging massive black holes as early as a month \textit{before} coalescence, extreme mass ratio inspirals, and ultra-compact binaries in the Milky Way \citep{LISAScience}. 
\end{itemize}

While these facilities have opened entirely new windows to study the universe, unlocking their full potential requires combining them with electromagnetic observations -- ``multi-messenger'' astronomy. For example, ground-based GW detections of stellar-mass compact binaries provide unique constraints on progenitor masses and spins, and exquisite tests of general relativity. Observations of the associated electromagnetic counterparts reveal the velocity and composition of associated outflows, as well as the host galaxy environment of the merger. Together, these systems can be used to constrain such fundamental (astro)physics questions as the equation of state of dense matter and the value of the Hubble constant.

Joint multi-messenger observations have already resulted in several spectacular successes in recent years. The joint discovery of GWs and electromagnetic radiation from the binary neutron star merger GW170817 was a watershed moment for astrophysics -- it confirmed that compact binary mergers can indeed power short gamma-ray bursts, that such systems are important (if not dominant) sites of heavy element nucleosynthesis, and that GWs travel at the speed of light. Similarly, the association of a high-energy neutrino with the blazar TXS 0506+056 demonstrated that such systems can indeed produce high-energy cosmic rays, and constrained the emission processes responsible for particle acceleration in these systems.

Despite these prominent successes, MMA/TDA observations present a unique set of technical (and sociological) challenges that have limited the pace of progress in recent years. Successful observations require coordination across myriad boundaries -- different cosmic messengers, ground vs.~space, international borders, etc.\ -- all for sources whose sky position may not be well known, and whose brightness may be changing rapidly with time. Lack of such coordination has at times had a profound impact -- for example, the poor localization of the second binary neutron star merger -- GW190425 -- resulted in highly redundant usage of finite optical follow-up, with some areas of the GW localization imaged more than 100 times, and others missed entirely. 

To help the MMA/TDA field realize its full potential, NASA and the NSF have convened a series of workshops to identify community needs and solutions to address them. At the first of these workshops in Annapolis, MD in August 2022, MMA/TDA \textit{infrastructure} -- hardware and software to enable coordinated observations -- was identified as key bottleneck for the field going forward. Motivated by this, the NSF sponsored a second workshop in Tucson, AZ in October 2023 titled ``Windows on the Universe: Establishing the Infrastructure for a Collaborative Multi-messenger Ecosystem.'' This resulting white paper is the outcome of that workshop.

\chapter{MMA/TDA Infrastructure Challenges and Proposed Solutions: Hardware}\label{sec:Hardware} 
\section{EM Messengers: Radio}
\label{sec:HW_EM_radio}
\noindent
{\bf Lead: Alessandra Corsi\\
Contributors: Poonam Chandra \\
}

\noindent \textbf{Radio Follow Up of Compact Binary Mergers---}In the context of multi-messenger astronomy, radio emission is particularly well-suited to uncover the most relativistic stellar explosions, involving core-collapses and compact object mergers that can be observed as long and short gamma-ray bursts (GRBs). Short GRBs, in particular, are sources of great interest.  Gravitational waves and electromagnetic observations at all wavelengths from GW170817 have revolutionized the field of gravitational wave astronomy \citep[see ][and references therein]{2017PhRvL.119p1101A,2017ApJ...848L..12A}. In the near-term future of LIGO, the radio band will enable follow-up observations that are critical to establish the presence of jets, probe their geometry and energy-speed structure \citep[e.g.,][]{Alexander2017,Hallinan2017,Corsi2018,Lazzati2018,Mooley2018a,Mooley2018b}, potentially constraining their sizes through Very Long Baseline Interferometry \citep[VLBI;][]{Mooley2018,Ghirlanda2019}. Radio observations can also shed light on the properties of the surroundings (ISM density), and can constrain particle acceleration mechanisms \citep[e.g.,][]{Lazzati2018,Hajela2019,Balasubramanian2021}. 

Generally speaking, radio observations are promising for uncovering off-axis compact binary mergers (as demonstrated by the case of GW170817), as well as mergers where a kilonova may not be detectable \citep[such as in neutron star - black hole mergers; e.g.,][]{Boersma2022}. While radio follow-up observations of GW170817 have provided us very high-impact results and greatly improved our understanding of relativistic jets, key questions remain to be answered. What is the diversity of ejecta outcomes in compact binary mergers involving at least one neutron star? How do the properties of the merger remnants map onto those of late-time radio flares? What constraints can we put on the equation of state of nuclear matter and on standard-siren cosmology using joint gravitational waves and radio detections?  The radio band is essential to answer these questions. Moreover, in the near future we will have the opportunity to clarify the potential link between fast radio bursts (FRBs) and compact binary mergers. Indeed, a $2.8\sigma$ association between GW190425 and FRB\,20190425A has recently been found \citep{Moroianu2023}. If this association is confirmed, FRBs could represent a new class of multi-messenger transients, associated with fast and coherent radio counterparts (as opposed to slower, non-coherent synchtron emission).

\noindent \textbf{Enabling New MMA/TDA Discoveries with Radio Observations---}Radio transients associated with mildly-relativistic massive star collapses or circumstellar medium (CSM)-interacting core-collapse supernovae, super-luminous supernovae (SLSNe), and fast blue optical transients, are also of interest in the context of multi-messenger astronomy, though multi-messenger detections have not been achieved yet. Indeed, transients with magnetars as compact remnants, as well as transients surrounded by dense CSM, are favorable sites for both non-thermal radio emission and neutrino emission \citep[e.g.,][and references therein]{Murase2019,Guarini2023}. 
IceCube Gen2 \citep{2021JPhG...48f0501A} and the next generation Very Large Array \citep[ngVLA;][]{Murphy2018}, are ideally suited to explore the link between neutrino and radio emission in these types of explosions. 
Another kind of supernova explosions, thermonuclear supernovae of type Ia, and their progenitors, are of interest in the context of multi-messenger astrophysics, and potential targets for the ngVLA, LISA, and deci-Hertz gravitational wave observatories \citep{Zou2020,Kinugawa2022}. 
In supernovae of type Ia, the explosion arises from a degenerate white dwarf star destabilized by mass accretion from a companion star, but the nature of their companion star remains poorly understood. 
Because a non-degenerate companion star is expected to lose material through winds or binary interaction before explosion, a supernova ejecta crashing into this material should result in radio synchrotron emission. While as of today we know of only one thermonuclear supernova for which radio emission has been detected \citep{Kool2023}, the order of magnitude sensitivity increase of the ngVLA (compared to that of the current VLA) could enable detection of  several such sources. 

Radio emission associated with massive and supermassive black holes could extend the reach of multi-messenger astronomy to the largest mass scales. We have already seen a neutrino from TXS\,0506+056 blazar \citep{2018Sci...361..147I}, the M77 nucleus \citep{2022Sci...378..538I}, and neutrinos from tidal disruption events \citep[TDEs; e.g.,][]{2021NatAs...5..510S}. High resolution VLBI and radio polarimetry observations are critical for AGNs, and the radio evolution gives unprecedented insights into the working mechanism of TDEs \citep{Horesh2021}. There is also unprecedented scientific opportunity in the synergy between PTAs, LISA and the ngVLA. PTAs are sensitive to gravitational waves from supermassive binary black holes at sub-parsec separations, when their orbital periods range from months to decades. Observations of these sources can be used to constrain the mass distribution of supermassive black holes and the galaxy merger rate, as well as shed light on the environments around them and the mechanisms that drive the merger. Gravitational waves from massive black hole binaries observed by LISA will constrain
the black hole masses and merger rates in a direct way, something currently inaccessible. 
On the other hand, the ngVLA unprecedented sensitivity and resolution could shed light on the life-cycle of massive black hole binaries in earlier stages of their evolution (1-100\,pc separations). Indeed, if one or both black holes in a pair are actively accreting (producing an AGN), multi-wavelength emission (including high-resolution VLBI observations) can directly mark their presence and locations \citep{2011MNRAS.410.2113B,2023arXiv230903252W}. Moreover, radio jets may form right before, during, and after the merger phase 
\citep{Mangiagli2022}.

\noindent\textbf{Radio Band Recommendations---}Overall, the National Radio Astronomy Observatory (NRAO) and its facilities have already played an outstanding role in enabling radio observations for multi-messenger astronomy, with facilities like the VLA, the GBO, and ALMA, proving critical in this field. 
In the near-term to mid-term future, current and planned radio surveys and facilities including CHIME, FAST, VLASS, SKA, DSA-2000 etc. will open the way to increased rates of transient discoveries. This will be paired with orders of magnitude increases in transient detection rates at other wavelengths enabled by upcoming focused survey by facilities such as Vera Rubin, Roman, Euclid, SVOM, ULTRASAT, etc. Moreover, an order of magnitude increase in well-localized compact binary mergers is expected with each future observing run of the LIGO detectors \citep{Petrov22}. Hence, the operations of National Radio Facilities need to rise up to this challenge. To this end, it is essential that:
\begin{itemize}
\item Improved radio arrays such as ngVLA are built to provide the sensitivity and resolution needed to extend the reach of radio follow-up observations to the larger horizon of distances of LIGO O5 (A+), A\#, and of next generation gravitational-wave observatories \citep{Evans2023}. We stress that the last would enable superb localizations of nearby compact binary mergers, opening the way to the discovery of radio counterparts independently of optical localizations.
\item The community continues to have access to VLA-like capabilities for gravitational wave follow-up in the transition between the VLA and the ngVLA era.
\item The NRAO works with the community to design strategies aimed at identifying high significance alerts and at optimizing the response to an increasing rate of gravitational-wave detections (especially with the VLA and ALMA). From present to future, this includes streamlining the creation of scheduling blocks, providing science ready data products on short timescales, as well as optimizing the use of subarrays for transient observations with the ngVLA. We note that a step forward in this respect is represented by the NRAO/GBO Transient Advisory Group, which has been set up to optimize transient science with the NRAO and GBO telescopes.
\end{itemize}

\section{EM Messengers: Millimeter}
\label{sec:HW_EM_mm}
\noindent
{\bf Lead: Joaquin Vieira\\
Contributors: Abigail Vieregg \\
}

A new generation of ground-based cosmic microwave background (CMB) surveys at millimeter wavelengths (mm-wave) are systematically mapping large portions of sky providing observations with arcminute-scale resolution.  As a result of their high spatial resolution, regular cadence, and unprecedented survey speeds, these surveys provide a new window on the largely unexplored mm-wave transient and variable sky.

Recently the capabilities of CMB experiments such as the Atacama Cosmology Telescope (ACT) and the South Pole Telescope (SPT) have been expanded to include time-domain science, with the first clear untargeted detections of mm-wave flaring stars \citep{guns2021,naess2021,tandoi2024}, extragalactic variable sources \citep{guns2021,hood2023}, and serendipitous measurements of the thermal emission from asteroids \citep{chichura2022}.

The mm-wave sky is expected to have a wide variety of detectable time-variable sources \citep{holder19,eftekhari2022}, including asteroids, flare stars, cataclysmic variables, x-ray binaries, nearby supernovae, tidal disruption events, gamma-ray bursts, and varying AGN. There is a demonstrated correlation between mm-wave and gamma-ray emission in blazars \citep{zhang2022}, pointing to future 
MMA synergies between the mm-wave and high-energy neutrinos as well as gravitational waves from binary neutron star mergers. 

CMB-S4 is the next-generation ground-based CMB experiment that will be online in the late 2020s. CMB-S4 is being designed with one of four primary science goals being the mm-wave transient sky. It will generate daily maps of roughly half the sky at mm wavelengths \citep{s42019}. 
Moreover, CMB-S4 was strongly recommended by the Astro2020 decadal survey \citep{decadal20}, with emphasis on the importance of CMB-S4 observations of the mm-wave transient sky, and CMB-S4 is the top recommended new major project in the 2023 P5 report \citep{p5Final}.
It is clear that CMB-S4 will have many transient events, but this is a new frontier in astronomy and likely to have many more surprises in the coming years. 

\textbf{Recommendation:} The next-generation of CMB experiments should plan to make data available to the broader astronomical community. This includes making transient alerts on timescales commensurate with the other observatories ($<$1 hour time scales), making calibrated maps and catalogs available one reasonable time scales ($\sim$1 year time scales), and making easily accessible data archives to facilitate MMA studies.
\section{EM Messengers: Optical and Infrared}
\label{sec:HW_EM_OIR}

\noindent
{\bf Lead: Tom Matheson\\
Contributors: Shenming Fu, Eric Hooper, Jayadev Rajagopal, Monika Soraisam, Ryan Lau, Armin Rest
}\\

The hardware needs for optical/infrared observations of
electromagnetic counterparts to multi-messenger events can be split
into two different phases: discovery and follow up.  Each of these
phases requires its own set of capabilities and techniques, although
there can be some overlap in facilities.  The success of the campaign
to find the kilonova associated with GW170817
\citep{2017ApJ...848L..12A}, as well as the extensive follow-up
observations \citep[e.g.,][and references therein]{MarguttiARAA},
demonstrates that the community has effective ideas about the
facilities needed to study multi-messenger events.  In addition, to
truly democratize access to the study of multi-messenger astronomy,
publicly available facilities should be a priority.

\subsection{Discovery}
Multi-messenger detections are poorly localized, with 90\% probability
regions on the sky ranging from 10s to 10s of thousands of square
degrees.  The increasing sensitivity of GW observatories allows
detections at greater distances, with subsequently weaker
signal-to-noise ratios, implying that the median 90\% credible area
stays relatively constant over runs O3, O4, and O5 \citep[$\sim1000$ square
degrees for binary neutron stars][]{Petrov22}.  It is likely that astronomers will focus on nearby events where gamma-ray and x-ray localizations are available, so that the search area will be 10s of square degrees (28 square degrees for GW170817), and thus within the capacity of current wide-field instrument.

As GW detections also provide a luminosity distance,
an alternative strategy is to search where the bulk of the stellar
mass might be at that distance--galaxies
\citep[e.g.,][]{gehrels16}.   A catalog developed specifically for
this purpose is the Extended Galaxy List for the Advanced Detector Era catalog
\citep[GLADE+;][]{Dalya22}.  The issue with this technique is that the
completeness of galaxy catalogs drops off rapidly, from ~60\% at
100Mpc to ~35\% at 500Mpc \citep[see Figure 2 of][]{Dalya22}.  It is
clear that galaxy targeting is effective, but this technique will not
work for the majority of detections, especially as the horizon for
multi-messenger detectors increases.  Wide-field, optical and infrared
imaging capabilities will be necessary to sustain multi-messenger science. 

\noindent \textbf{Augmenting Existing Capabilities---}Adding new capabilities without supporting existing ones does not
solve the problem, it merely shifts it.  The Dark Energy Camera
(DECam) on the Blanco 4m telescope at CTIO, with a 3 square degree
field of view, has already demonstrated its effectiveness at finding
counterparts to GW events with its independent discovery of the
kilonova associated with GW170817 \citep{soaressantos17}.  There are few wide-field
telescopes with significant aperture that can image the southern
sky--and none with public access for the US community.  When the Rubin
Observatory comes online in 2025, its 10 square degree field of view
will surpass DECam's reach, but only a small ($\sim3\%$) fraction of its
time will be available for multi-messenger searches.  DECam would
continue to provide a larger fraction of time for discovery and
identification of counterparts and it could be complementary to Rubin,
searching different areas or matching Rubin and using a different
filter.  At a minimum, the unique capability that DECam provides
should be maintained.

Enhancements to improve the capability of DECam for multi-messenger
science are possible.  These are not hardware solutions, \emph{per
  se}, but are directly tied to DECam itself.  NOIRLab is currently
adapting the Rubin data processing pipeline to operate on DECam data
and thus provide rapid image subtraction.
Discussions between NOIRLab Community Science \& Data Center (CSDC) and Rubin data management are also taking place to potentially develop joint processing directly at the Rubin US Data Facility.
Support for the NOIRLab program,
including ongoing improvements to a database of templates for
subtraction and operational costs, would directly benefit the US MMA/TDA
community, freeing them from the data processing effort and cost.  In
addition, both time allocation and scheduling to accommodate ToOs,
including interrupts to classically scheduled nights should be
implemented.  Further integration into the AEON system is important.
This may require overhead in the form of additional user support and
queue operation.

The NEWFIRM instrument has recently returned to the Blanco telescope.
While it has a smaller field of view than the optical instruments, it
is large for an infrared instrument.  Enhancements for NEWFIRM could
include a real-time data reduction system, as well as the allocation
and scheduling changes that would help DECam.

There are also several wide-field time domain surveys on smaller aperture telescopes that are effective tools for wide-field searches when the detections are likely to be within its magnitude range, in particular when the search area spans thousands of degrees. ZTF, ATLAS, PS1, ASAS-SN, and a number of future surveys like BlackGEM and the La Silla Southern Sky Survey make up for their shallower depths with their FOVs in the tens of degrees. Future funding for these surveys is often uncertain or only guaranteed for a small number of years. Supporting the ongoing operations of at least some of these surveys would be valuable, especially \textit{with firm commitments for immediate public access to images and alerts (not just the vetted candidates)} when it is engaged in MMA/TDA searches.

\noindent \textbf{New Capabilities---}There are several wide-field, large-aperture telescopes that can image
the northern sky, but public access for the US community is extremely
limited in time and capability.  With the growing sensitivity of 
non-EM detectors, the greater capacity means greater distance, with most of
the accessible volume at these larger distances.  We need telescopes
with enough aperture to detect these distant sources.  In addition,
larger facilities can reach depth quickly enough that they can image
fields in more than one filter.  Color can be a significant
discriminant in identifying objects such as kilonovae.  An entirely
new large telescope could be cost-prohibitive, but the WIYN 3.5m
telescope at Kitt Peak exists and has the framework of an instrument
capable of providing a 1 square degree field of view in the One Degree
Imager (ODI), thus providing a relatively inexpensive and shorter term
path to provide this capability.  WIYN is at an excellent seeing site
and, incorporating the pipeline and operation solutions described
above for DECam, a refurbished ODI with a new detector array would
provide the US community with exactly the northern instrument that is
currently missing from the suite of tools available.

While support for small-aperture surveys like ZTF and ATLAS can fulfill the need for relatively bright, long-lived sources, supporting it should not prevent development and support of new, innovative surveys and instruments.
Development of new hardware and technology that enable faster survey speeds at higher cadence and deeper sensitivities are crucial for exploring the physics that can drive the rapid $(\lesssim$ hr) variability of EM counterparts to GW events. An example of such a project in development is the Argus Optical Array \citep{Law2022}, which will 
survey 20\% of the entire sky every second. 
Therefore, in addition to new surveys with existing instruments, new instruments that can push the depth-cadence phase space to enable groundbreaking science on EM counterparts of multi-messenger events, should be supported. 
Again, it is important to emphasize that {\bf public access is a vital criterion for any new survey}. 

Although imaging surveys have primarily focused on optical wavelengths,wide-field near-infrared (IR; $\sim1-2$ $\mu$m) imaging capabilities present an important resource for follow-up observations given that IR emission from a kilonova is expected to be isotropic, long-lived ($>1$ week), and ubiquitous \citep{Kasen2017}. Notably, based on kilonova model predictions, the detection rates in the near-IR could be up to a factor of $\sim10$ higher than in optical wavelengths \citep{Zhu2021}. A limiting factor for producing wide-field near-IR imaging instruments has been the cost-per-pixel of IR detectors. However, the recent commissioning of the Wide-field Infrared Transient Explorer (WINTER; \citep{Lourie2020,Frostig2022}) has demonstrated that indium-gallium-arsenide (InGaAs) detectors, which are $\sim1/3$ the cost per pixel of the more commonly used mercury-cadmium-telluride (HgCdTe) detectors and do not require cryogenic cooling, provide a more affordable alternative to build wide-field near-IR instrumentation.

\subsection{Follow Up}
Follow-up observations of counterparts to MMA/TDA events could potentially use every available optical and infrared instrument, both ground- and space-based, that could achieve the signal-to-noise ratio necessary to be scientifically useful.  In practice, because we expect optical/IR countrparts to be faint, follow-up is typically photometry and (low-resolution) spectroscopy.   For such faint targets, the sensitivity of space-based observatories like the \textit{Hubble Space Telescope} (HST) and the \textit{James Webb Space Telescope} (JWST) are crucial for conducting spectroscopic and photometric follow-up observations. However, due to constraints on observing time and scheduling, follow-up observations with HST and JWST need to be carefully planned with exposure times tuned to the brightness of the optical/IR counterpart. It is therefore important to coordinate ground- and space-based observations to ensure rapid follow up with sufficiently high signal-to-noise ratio detections.

For brighter targets like GW170817, we should be prepared to support every capability than can be deployed. For general time-domain follow up, especially in the Rubin era, there is clearly a lack of resources. Although the current rate of multi-messenger events is relatively low, ensuring we have a suite of publicly accessible follow-up resources at optical/IR wavelengths is a high priority as the EM discovery rate increases with new survey facilities coming online in the decade. In some instances (e.g. Local Group SNe, FBOTs, short orbital period binaries) high time sampling and/or high spatial resolution optical and IR imaging capabilities will present unique science opportunities. These instruments exist today (e.g. 'Alopeke and Zorro) and therefore provide a low-cost, present-day capability that should be maintained.

\noindent \textbf{Augmenting Existing Capabilities---}Almost all the augmentation of existing optical and infrared hardware
would be focused on software, where the key enhancement would be efficiency. Time allocation and scheduling for rapid and sustained interrupts require policy-level decisions.
Effective distribution of observations would require support of AEON
and similar systems.  Rapid data reduction systems would enable quick
vetting of candidates.   Examples of how effective this can be are
readily apparent in the time-domain operations of SOAR and Gemini.

\subsection{Recommendations}

\begin{itemize}

\item Maintain DECam as a viable discovery instrument

\item Provide public wide-field imaging capability in the north (ODI)

  \item Support pipeline and data infrastructure to make DECam data
    more accessible

  \item Support NEWFIRM at the same level as DECam

  \item Sustain and extend ZTF as a survey and discovery instrument


  \item Support the development of new surveys, instruments, and telescopes enabling fast survey speeds, high cadence, and deep sensitivity 

  \item Support the development of wide-field IR imaging instruments

  \item Support software and policy improvements for telescope and
    instrument coordination and efficiency

\end{itemize}
\section{EM Messengers: Ultraviolet}
\label{sec:HW_EM_UV}
\noindent
{\bf Lead: Aaron Tohuvavohu\\
}
Multi-messenger source classes that produce gravitational wave and particle radiation detectable with current instruments are highly energetic. These sources include mergers, jets, stellar collapse, and other high curvature, acceleration, and density environments. Such energetic processes naturally heat the matter to high temperatures, thus thermally radiating in the ultraviolet (UV) before cooling. UV observations are therefore \textbf{essential} for early-time discovery, and characterization, of these multi-messenger sources in the EM sector. The required observations generically require some subset of wide-field of view, high sensitivity, spectroscopy at R$>1000$, and fast response ($<$ hour).

Due to the atmospheric opacity, the UV is uniquely accessible from space-based platforms, and our capacity has therefore been historically limited. UV observations of time-domain and multi-messenger source classes is in high demand, with the modest, but nimble, Swift UV/Optical Telescope (UVOT) performing the large majority of followup and characterization observations for the community, providing UV data for $\sim1000$ transients per year. However, Swift/UVOT is insufficient for many required measurements, due to its narrow field, limited sensitivity, and low-resolution grism, which limit our ability to fully exploit the UV domain for discovery, followup, and characterization respectively. Hubble provides the community with significantly enhanced UV sensitivity and spectroscopy, but its slow operations and scarce availability leave a substantial science gap.

ULTRASAT (launch 2027), an Israel-DESY-NASA mission, will be an incredibly powerful discovery engine in the UV time-domain. ULTRASAT will close one of the aforementioned gaps, with its extremely large field of view ($\sim200$ deg$^2$) providing very fast survey speed for discovery of UV transients in the near-UV (220-280 nm) band with reporting latency of $<15$ minutes. However, the ULTRASAT revolution will be incomplete without the necessary instruments to follow-up and characterize the sources discovered. 
This will require more sensitive UV instrument(s), with coverage extending through the far UV, multiple photometric bandpasses (simultaneous for rapidly evolving sources), spectroscopy (R$>1000$), and agile ToO driven operations for high efficiency follow-up ($<$ 1 hour).
NASA should prioritize a \textbf{ToO-driven mission}(s) with the above capabilities in order to access the extremely compelling physics uniquely encoded in the early, hot, environments of energetic multi-messenger sources. 
\section{EM Messengers: X-ray \& Gamma-ray}
\label{sec:HW_EM_Xray}
\noindent
{\bf Lead: Adam Goldstein\\
Contributors: Nick White, Aaron Tohuvavohu, Eric Burns, Jamie Kennea, Hugo Ayala
}\\

\subsection{X-ray}
The X-ray sky has a relatively low incidence of transient activity compared to optical wavelengths, therefore historically X-ray detections have been an powerful discriminator to identify and localize a transient. Although upcoming optical surveys like Rubin and the existing ZTF can localize orphan afterglows, X-ray observations of these candidates can easily confirm the GRB nature. This process has been performed and confirmed many times with the X-ray Telescope onboard the Neil Gehrels Swift Observatory. In addition for Neutrino searches, Swift was the first telescope to identify TXS~0506+056 as a possible counterpart, utilizing rapid follow-up observations of the IceCube error region. 
At longer timescales, in the case of GW 170817, late time X-ray observations with the Chandra observatory provided strong limits on the jet physics in the system. X-ray observations are space unique, and in high demand thanks to the current open access to target-of-opportunities at multiple missions, but exemplified by the workhorse Swift, which accepts ~1700 TOOs/years. In the near future, many current X-ray observatories, including Swift and Chandra may end. Therefore there is a strong need for wide-field/all-sky workhorse X-ray telescopes that can provide arcminute-scale localization of transients toreplace these vital components in the MMA/TDA infrastructure to accomplish new science as other observatories improve their coverage and sensitivity. As Swift has clearly demonstrated, the need for rapid response, high efficiency observing has strong community support, and NASA should prioritize the replacement of Swift, as outlined by the Astro2020 Decadal Survey. A next generation X-ray telescope combining Swift's rapid response capabilities to follow-up GRBs in minutes or seconds, not hours, combined with higher sensitivities could cover the three science cases briefly outlined above, and much more.

\subsection{Ground-based VHE Gamma ray}
Although telescopes like Fermi-LAT can detect photons above 100s of GeV, ground-based telescopes like Imaging Air Cherenkov Telescopes (IACTs) or particle detector arrays are required to detect gamma rays at a high enough rate and with enough statistics to become significant observations. This is because space telescopes are limited in size due to the requirements of spacecrafts. VHE gamma-ray astronomy provides unique insights into some of the most energetic and extreme phenomena in the cosmos. Most of the current experiments rely on the Cherenkov effect. Gamma rays, when they interact with the atmosphere, produce an extensive air shower of secondary particles composed mostly of lower-energy gamma rays and electrons. Electrons traveling in a refractive media can produce Cherenkov radiation which is the light that is collected by the ground-based detectors. Detectors like VERITAS, H.E.S.S., MAGIC, HAWC, LHAASO, CTA, etc. not only complement observations in other wavelengths, but also have made important contributions of their own. For example, ICATs have made discoveries of TeV sources that lack a counterpart in other wavelengths, and HAWC has found new objects like the TeV Halos. Furthermore, gamma-ray bursts have also been observed from the ground: GRB 180720B was detected by H.E.S.S; GRB 1901114C was detected by MAGIC in the TeV range; and GRB221009A was detected by LHAASO with observed photons up to tens of TeV. These discoveries have established VHE gamma-ray astronomy as a field of its own.
Ground-based experiments can also perform follow-up observations of multi-messenger sources like neutrino or gravitational-wave alerts. Extensive air shower survey experiments, such as HAWC and LHAASO, can monitor the sky due to their large field of view. This large sky coverage also makes them ideal for follow-up of the large uncertainty regions of gravitational wave events. IACTs, with better angular, energy resolution and sensitivity, can perform searches in specific parts of the sky. Given that more multi-messenger experiments, like IceCube Gen2, or gravitational wave runs like O5 are anticipated in the next 10 years, maintaining the funding for the ground-based gamma-ray facilities will be important for the MMA/TDA science.
The most effective multi-messenger and multi-wavelength program at very-high energies will consist of both particle detectors like HAWC/SWGO and IACTs like CTA. Such a program would combine the main strength of particle detectors to measure extended sources and conduct Galactic Science and the capabilities of IACTs to conduct extragalactic science and high angular resolution Galactic observations coincident with other high-resolution wavelengths. 

As a recommendation, future experiments like Cherenkvo Telescope Array (CTA) and the Southern Wide-field Gamma-ray Observatory (SWGO) are in development and will need support. Both were already recommended in the Astro2020 Decadal. 
In the case of SWGO, current designs consider an experiment larger than HAWC and at a higher altitude. The site selection for SWGO is in process and it will be built in a place in South America. This is important since no large-field-of-view survey ground-based gamma-ray observatory is located in the southern hemisphere. Being in the southern hemisphere will notably give us better access to the Galactic Center. Furthermore, its field of view will have an overlap with the current HAWC experiment, making it ideal to have a joint monitoring system of the sky in high-energy gamma rays. It will also complement observations done already by H.E.S.S. on the Galactic plane and have synergistic contributions with CTA South.
CTA has already received funding by NSF for more than 10 years and has already deployed working IACTs including the first Large-Sized Telescope (LST) telescope. Furthermore, the CTA Schwarzschild-Couder Telescope at the Whipple Observatory has already made measurements of the Crab Nebula. Stable and continued funding will help CTA with its scientific objectives.

\subsection{Space-based Gamma ray}
Hard X-ray and Gamma-ray observatories provide the first EM evidence that a relativistic jet was launched from a cataclysmic event: typically a massive core-collapse or a compact merger. These observatories collect important properties of the jet and central engine, such as the spectrum, energetics, and activity time.  They also provide critical pieces of supporting information to the community, such as the start time of the jet and the localization.  For more than 20 years, the community has used this information to follow-up high-energy transients with telescopes on the ground and in orbit, building up an entire ecosystem of partnerships and collaborations.  It is more important than ever to continue, and expand, this ecosystem. The capability to sensitively and continuously monitor the entire sky in gamma-rays is the most efficient way to detect counterparts to multi-messenger signals.  In fact, as International Gravitational-Wave Network (IGWN) becomes more sensitive, gamma-ray bursts (GRBs) will be the dominant counterpart associated with NS mergers, by far. The GRB localization, combined with the GW localization, further enables follow-up, and continues to be the most likely way to explore NS mergers as multi-messenger sources over the next decade and beyond. Swift and Fermi, the workhorse missions that have the capability to make these observations, are aging, are one critical failure away from end-of-life, and are unlikely to observing into the 2030s.  While there are international pursuits currently and in the near future, such as AstroSat and SVOM, these missions are largely less capable than Fermi and Swift, and they do not always have open data policies. Without focused replacement and/or upgrades of some of their key capabilities, many multi-messenger investigations will suffer, and some investigations may not be possible. Therefore, sensitive wide-field hard X-ray and gamma-ray missions with the \textbf{capability to localize transients to within a few arcminutes} needs to continue to be a priority for NASA astrophysics into the 2030s.  \\

\subsection{Space Communications}
MMA/TDA tends to place strong requirements on communications infrastructure for space based missions. GRB detectors need to communicate detections in realtime to the community, requiring an always on-communication capability. Previously NASA provided this using the Tracking and Data Relay Satellite System (TDRSS). However, for future missions NASA is not allowing the use of TDRSS as it winds down. Although commercial replacements are being studied, they have not yet been selected. Therefore NASA should prioritize filling this gap. In addition to reporting of transients, increasingly it is useful to perform rapid follow-up, requiring spacecraft be commanded to repoint in realtime. The need to perform real-time follow-up means that commanding be available either at short latency, or continuously. This can drive costs and also stresses the existing infrastructure, disadvantaging MMA/TDA missions in the eyes of selection panels. In July 2023 NASA ratified the MMA/TDA Communications Science Analysis Group to produce a report on this issue, which should be completed in early-mid 2024. It is recommended that NASA follows the recommendations of this SAG, in order to ensure that the communications infrastructure for future MMA/TDA missions best enables the science.

\subsection{Space Observatories as Infrastructure}
MMA/TDA requires a change in how NASA evaluates and selects space-based missions.  While NASA considers communications, archives, and public analysis tools as part of necessary infrastructure, space-based science missions also need to be considered as critical infrastructure.  Consider the scenario where a GW network detects NS mergers but there are no sensitive high-energy monitors operating. There is a significant loss in science in that scenario, because a high-energy monitor is the best and most efficient method to determine if a relativistic jet was launched by the merger. The loss of science in such a scenario even goes beyond that of astrophysics, as it even affects inferences on fundamental physics and theories that can only be tested in these extreme astrophysical scenarios.  Therefore, it is important for certain space-based missions to be considered as critical infrastructure to maintain and sustain critical observing capabilities as ground-based and other space-based facilities come online.

The InterPlanetary Network (IPN) is the premier example of space observatories as infrastructure. The United States has maintained continuous and complete coverage of the soft gamma-ray sky for more than half a century with the IPN. The data from wide-field monitors are collated, enabling precise localizations for gamma-ray transients. These combined observations and alerts are a foundational need in this new era of time-domain and multimessenger astronomy. It is necessary to launch new dedicated instruments and to integrate them into the IPN for the foreseeable future, with support at all levels to foster cross-divisional science at NASA. Intentional strategic support is needed to support the necessary effort and expertise, and to allow for long-term planning.

\subsection{Programmatic selection of X-ray/gamma-ray missions}
The top recommendation from the Astro2020 Decadal for sustaining activities in space was a Time Domain and Multi-messenger program to ``realize and sustain the suite of capabilities required to study transient phenomena and follow-up multi-messenger events."  This recommendation was ranked as a higher priority than even the Probe line of missions, which is currently being pursued by NASA instead of the time domain and multi-messenger program.  We strongly recommend that NASA abides by the recommendations of the Astro2020 Decadal, \emph{in order of priority}, and consider such a program.

This requires a change in how NASA currently considers and selects missions.  Missions below the scale of a flagship are considered as standalone investigations into a singular science case that is considered without regard to the thematic pursuit advised in the Decadal Survey.  This approach is wholly inadequate for MMA/TDA, which necessarily requires a strategically designed portfolio of instruments covering the EM spectruma and other messengers. For MMA/TDA to be successful, NASA needs to consider a strategic, holistic approach to MMA/TDA missions.

Another problem to NASA's approach regarding MMA/TDA, is that missions are typically considered as an all-in-one package, required to satisfy their entire proposed science with only their on-board instruments or by requiring written commitments guaranteeing observations from facilities years in advance, which may impose a financial cost on the proposed mission if it is even feasible for the facility to make such a commitment that far in the future. Many publicly funded observatories such as NOIRLab utilize a 6- to 12-month time allocation process. Space mission proposals typically can take 5-7 years to realize. There is no process for NOIRLab (and ESO) observatories to make the necessary advance commitment.  This is a barrier that has impacted several mission proposals, with solutions including purchasing telescope time and/or utilizing directors discretionary time, typically on private observatories. This excludes many large publicly funded telescopes. 
As an example, the MMA/TDA Gamow Explorer mission proposed to NASA in the 2021 MIDEX Explorer competition would have used gamma-ray burst (GRB) afterglows as bright backlights to probe the high-redshift reionization epoch. 
Rapid follow-up from large telescopes with NIR spectrographs to measure the damped Lyman alpha wing and metal absorption lines was an essential part of its science case. 
However, the limited amount of time commitments (none on large public facilities) that the team was able to secure for their proposal -- years in advance of the mission -- led to the NASA review panel judging it as ``high risk.'' 
Recommendations to address this challenge are: 
\begin{itemize}

\item {\bf NSF} should develop, in coordination with {\bf NASA}, a policy that supports future MMA/TDA space mission proposal observing time requirements in an equitable way.

\item {\bf NASA} treats external observatory commitments as a programmatic factor, not as a scientific or development risk -- the approach taken for international contributions. The programmatic risk would consider required follow-up based on feasibility and past performance of the proposing team. For existing NASA assets e.g., JWST, the observatory director in coordination with NASA HQ provides a mechanism to support MMA/TDA Explorer proposals time commitment requirements in an equitable way.
\end{itemize}

Furthermore, there is currently no standard way to evaluate how a mission's service or contribution as critical infrastructure augments, or even exceeds, its own self-contained science case. In order to for NASA space missions to effectively contribute to time-domain and multi-messenger astrophysics, {\bf we recommend a dedicated call for MMA/TDA-focused sustaining infrastructure missions.  Such a call could either be in place of one of the Explorer calls or augment an Explorer call to appropriately focus on MMA/TDA missions}. This is because the guidelines by which time-domain/multi-messenger infrastructure missions should be judged will certainly be different than standard Explorers due to the importance of sustaining capabilities and infrastructure.

\newcommand{\Asharp}{A\textsuperscript{$\sharp$}}
\section{Non-EM Messengers: Gravitational Waves}
\label{sec:HW_nonEM_GW}

\noindent
{\bf Lead: Patrick Brady\\
Contributors: Salvatore Vitale
}\\

With the first observation of gravitational waves from a binary black hole merger in 2015~\citep{GW150914}, LIGO and Virgo opened the field of gravitational-wave astronomy. Then in 2017, LIGO and Virgo detected gravitational waves from a pair of neutron stars~\citep{AbEA2017b}. A gamma-ray burst was observed by the Fermi-GBM instrument $\sim 1.7$ seconds after the merger time~\citep{AbEA2017e}. Within 24 hours, an optical transient AT 2017gfo was detected in NGC 4993 leading to an unprecedented follow-up observation campaign~\citep{2017ApJ...848L..12A} that stretched over more than a year after the event.

\begin{figure}
    \centering
    \includegraphics[width=0.9\linewidth]{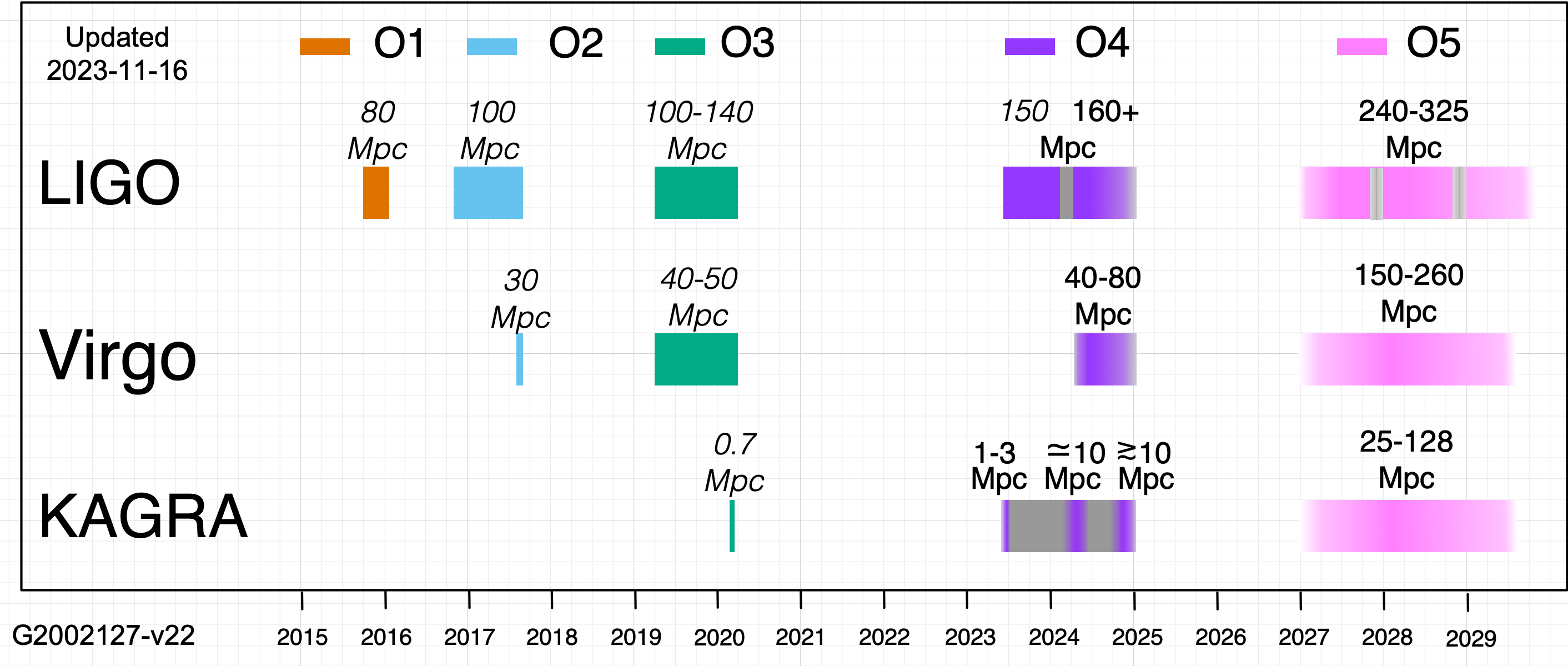}
    {\caption{\label{fig:lvk-timeline} The LIGO-Virgo-KAGRA observing plans through 2029. Observing runs are labelled O1, O2, etc. The distances above each bar represent the binary-neutron-star inspiral detection range described in the text. The A+ project will bring the LIGO detectors to their ultimate O5 sensitivity. Virgo is currently reconsidering its plans for O5 in view of difficulties achieving its O4 sensitivity goal. LIGO, Virgo and KAGRA anticipate observing into the next decade to dovetail their operations with next-generation detectors Cosmic Explorer and Einstein Telescope. Figure Credit: LIGO-Virgo-KAGRA Collaboration}}
\end{figure}

To achieve the sensitivity needed to detect gravitational waves from astrophysical sources, gravitational-wave detectors require many low-noise components and subsystems to be brought together and controlled with the utmost precision~\citep{PhysRevLett.116.131103}. The gravitational-wave frequency window for earth-based, interferometric detectors is bounded below at around 10Hz by seismic noise while laser shot noise dominates at high frequencies leading to a window of a few kHz. Many other fundamental and technical noise sources also limit the sensitivity in the mid-frequency range. Detector sensitivity is often summarized in the binary-neutron-star (BNS) range $\mathcal{R}_{\mathrm{BNS}}$ which is the average distance at which BNS inspiral would produce a matched filtering signal to noise ratio of 8; the averaging is over sky location and inclination angle. The distance at which an optimally oriented and located BNS inspiral would produce a signal-to-noise ratio of 8 is called the horizon distance~\cite{Chen_2021}; it is about $\sqrt{5} \mathcal{R}_{\mathrm{BNS}}$. The Advanced LIGO construction project started in 2010 after almost a decade of planning. Observation runs have been interspersed with periods of detector improvement since 2015 as shown in Fig.~\ref{fig:lvk-timeline}. 

To date, more than 100 binary black hole mergers, 2 binary neutron star mergers and a handful of neutron-star-black-hole mergers have been detected. The fourth LIGO-Virgo-KAGRA observing run (O4) started on 24 May 2023 and is expected to continue for 20 months including two months of commissioning slated for the first quarter of 2024. Compact binary mergers are being detected once every few days. After 6 months, there are a couple of NSBH and no BNS candidates~\cite{gracedb}. These low rates reinforce the importance of continuing to improve the sensitivity of ground-based gravitational-wave detectors.

Over the next 5 years, work on the LIGO detectors should allow them to reach $240 \textrm{Mpc} < \mathcal{R}_{\mathrm{BNS}} < 325 \textrm{Mpc}$, by way of the A+ upgrade project, thus increasing the rate of detected binary mergers by up to a factor of $\sim 8$ by the end of the 2020s. The detection rate of BNS mergers has substantial uncertainties arising from small-number of detections to date and the anticipated sensitivity range leading to estimates from 0.5 to 3 per month.

Ground-based gravitational-wave detectors are a cornerstone of MMA/TDA. It is important to continue to improve the current detectors, including LIGO, and to begin construction of next generation detectors such as Cosmic Explorer and Einstein Telescope. 
The LIGO Scientific Collaboration is studying a set of upgrades called \Asharp{} \citep{ligo_postO5_report}{} that could increase the detection rate of BNS by ~5 beyond A+ and could be operational early in the 2030s. LIGO-India should also begin operating around the same time. Both Virgo and KAGRA are also considering upgrade paths to boost their sensitivities. Operations of three upgraded LIGO detectors, Virgo and KAGRA would dovetail with operations of next generation facilities.
The US-led Cosmic Explorer~\citep{Evans2023} and European Einstein Telescope~\citep{Punturo_2010} projects promise to take gravitational-wave astronomy to the next level. Cosmic Explorer is conceived to be located above ground with arm lengths of 40 km thus increasing their BNS range by at least a factor of 10 over A+. These detectors will detect BBH mergers throughout the Universe and would increase the detection rate of BNS mergers to many per day. They will enable exquisitely accurate measurements of gravitational-wave source parameters delivering breakthroughs in our understanding of how compact object formation evolved throughout cosmic history, as galaxies were formed and stars became more metal-rich, the internal structure of neutron stars, cosmology and test the limits of current gravitational theories.

Over the next five years, we need to develop and refine a full suite of instrumentation and cyberinfrastructure to support the discovery, classification, and follow-up of multi-messenger sources. The importance and power of this approach is exemplified by GW170817. With appropriate investment, collaboration and coordination, NASA and NSF can enable an unprecedented era of multimessenger astronomy with ground-based gravitational-wave detectors as the cornerstone.

Pulsar timing arrays are delivering access to nHZ gravitational waves from the mergers of supermassive black holes. This approach to gravitational-wave observations relies on the availability of radio telescopes and support of programs to time an array of pulsars across the galaxy. Pulsar timing arrays require regular monitoring of many dozens of pulsars over decades in order to reach nHz GW frequencies. The sensitivity improves with the number of pulsars, and as the number of pulsars increases so does the amount of telescope time required to observe them all. See Sec~\ref{sec:HW_EM_radio} for a description of important radio-astronomy capabilities. We also note that accurately-timed millisecond pulsars are target sources of gravitational-wave emission in the  ground-based detector band. 

The space-based Laser Interferometer Space Antenna (LISA), slated for launch in the mid 2030s, will detect gravitational waves between the 0.1 mHz and 1 Hz. The mission has a nominal 4 year lifetime and may be extended up to a decade. With unique science goals of its own, LISA will also enable both multiwavelength observations of some binary-black systems and multimessenger observations of systems such as white dwarf binaries in our galaxy. 

\section{Non-EM Messengers: Neutrinos and High Energy Particles}
\label{sec:HW_nonEM_neutrino}

\noindent
{\bf Lead: Lu Lu\\
Contributors: Erik Blaufuss, Marcos Santander, Tianlu Yuan}\\

The observation of high-energy particles, in particular astrophysical neutrinos from distant sources, can provide a unique view deep into the most powerful astrophysical objects in the Universe. The location of these cosmic accelerators, thought to give rise to the observed flux of cosmic rays \cite{Coleman:2022abf}, remains a mystery more than 100 years after their discovery.  Neutrinos, electrically neutral and travelling near the speed of light, can escape these dense particle acceleration regions due to their small interaction cross sections and, once detected on Earth, accurately point the way back to their accelerators.  Such dense and energetic acceleration regions are expected to be visible across the electromagnetic spectrum, making them likely multi-messenger sources. Broadly, there are many areas within MMA/TDA where astrophysical neutrinos can play a crucial role. These will take coordination and collaboration as future GW, EM and neutrino observatories come online. Establishing supportive infrastructures for the real-time dissemination of information will be crucial. Within this framework, two primary areas of focus emerge: 
\begin{itemize}
    \item Integration and coordination between neutrino observatories. With IceCube in continuous operation and several neutrino telescopes either planned or in construction stages ~\citep{KM3Net:2016zxf,GRAND:2018iaj,P-ONE:2020ljt,Romero-Wolf:2020pzh,IceCube-Gen2:2020qha,POEMMA:2020ykm,PUEO:2020bnn,RNO-G:2020rmc,Ye:2022vbk,Malyshkin:2023qnc}, establishing a global network of neutrino observatories becomes imperative.
    \item Collaboration and synergies among neutrino observatories within the MMA /TDA framework including GW and EM.
\end{itemize}

Terrestrial observation of high-energy neutrinos at Earth is not without challenges.  Construction of IceCube required instrumentation of a cubic kilometer of Antarctic glacial ice at the South Pole.  This glacial ice has complex optical properties that impact the Cherenkov light propagation from charged relativistic particles produced in neutrino interactions~\citep{tc-2022-174} and impact the directional uncertainty found when reconstructing neutrino directions.  The observed angular uncertainties of neutrino events are key drivers of the sensitivity of IceCube to searches for neutrino emission from point sources~\citep{IceCube:2022der}, where improved angular uncertainty would reduce the impact of atmospheric backgrounds and make IceCube sensitive to smaller neutrino fluxes.  

IceCube issues alerts to the astrophysical community following the detection of a neutrino likely of astrophysical origin~\citep{IceCube:2023agq}. The angular direction and  uncertainty of these single events is calculated and can be strongly impacted by geometry of the event in the detector, the morphology of the event, and the limits of our understanding of the glacial ice properties on light propagation. In certain instances, 90\% error regions can exceed 3 sq.~deg.~for muon-neutrino induced track alerts, leading to source confusion in identification of potential counterpart objects. For cascades, induced by particle showers from other neutrino flavors or neutral current interactions, the error regions are typically much larger. 
A challenge in obtaining accurate directional reconstruction is in the accurate modeling of particle showers and the Cherenkov light they produce. A good understanding of ice and instrument is crucial. Current \emph{in-situ} calibration devices have performed well in refining our understanding of large-scale features in the glacial ice~\citep{tc-2022-174}. However, a better understanding of local ice properties near each sensor, as well as their orientations and  positions is needed to exploit IceCube event signatures in full. Another challenge in the domain of MMA/TDA neutrino astronomy relates to the signal event rates. Even at cubic-kilometer volume, the observed fluxes of neutrinos above atmospheric fluxes combined with the small neutrino cross section yield small numbers of individually high-confidence astrophysical neutrinos.  IceCube currently identifies $\sim$6 Gold-level track alert events per year over background~\citep{IceCube:2023agq}.

To address these challenges, various strategies are being employed. One significant initiative, the IceCube Upgrade, is scheduled for deployment in 2026~\citep{Ishihara:2019aao} and aims to enhance the observatory's sensitivity to GeV-scale neutrinos. This upgrade will involve the installation of seven new strings, housing approximately 700 multi-pixel optical sensors. Additionally, self-contained light-emitting devices such as the POCAM, the Pencil Beam, and acoustic emitters will be deployed to effectively calibrate the properties of the glacial ice and conduct in-situ evaluations of detector responses. The calibration program aims to achieve a cascade angular reconstruction closer to the statistical limit, with the goal of improving the angular resolution by over a factor of two compared to the existing systematic limit observed with the current IceCube configuration. With these better calibrations in hand, the 15-year archive of IceCube neutrinos can be reexamined with improved sensitivity, as well as improving the future astrophysical neutrino community alerts.

However, the detection of a larger number of astrophysical neutrinos necessitates new instrumentation. Several new neutrino observatories are either under construction or being considered. The KM3Net~\citep{KM3Net:2016zxf} ARCA detector, currently under construction, will instrument a volume similar to IceCube in the Mediterranean Sea, providing a complementary view of the sky compared to IceCube's southern location. At higher neutrino energies, the Pierre Auger Observatory is currently undergoing an upgrade~\citep{Stasielak:2021hjm} to enhance detection capabilities for neutral particles, including neutrinos at beyond $10\,\mathrm{EeV}$. Furthermore, the RNO-G detector~\citep{RNO-G:2020rmc}, currently in the construction phase, operates on the Askaryan effect, designed to capture radio signals from neutrinos above 30 PeV. This pioneering effort provides crucial initial measurements in this energy range and contributes significantly to the technology development for future detectors. Ultimately, a larger in-ice optical and radio neutrino detector is imperative to significantly increase the number of observed astrophysical neutrinos. The IceCube-Gen2 detector \cite{IceCube-Gen2:2020qha} aims to achieve this by expanding its instrumented volume to nearly an order of magnitude larger than IceCube. Additionally, it incorporates a 500\,$\mathrm{km^{2}}$ radio array specifically designed to resolve the high-energy neutrino sky across a spectrum ranging from $10\,\mathrm{PeV}$ to EeV energies. In the 2023 report by the Particle Physics Project Prioritization Panel (P5)~\citep{p5Final}, a recommendation was made to maintain continuous coordination and planning between NSF-OPP and projects like IceCube-Gen2 at the South Pole. This initiative aims to sustain NSF's infrastructure at the South Pole, ensuring the facilitation of significant scientific discoveries in the coming decade. 

\chapter{MMA/TDA Infrastructure Challenges and Proposed Solutions: Software}\label{sec:Software}
Software development is one of the crucial tasks to keep the ecosystem of multi-messenger astrophysics running. Alerts need to go out at low latency, and are often acted upon immediately by the astronomical community, who need to coordinate and make real-time decisions on what candidate counterparts need to be pursued.  Each of these components (and others discussed below) require robust, inter-operable software. 
 Some of these software suites are operated at the `PI level' while others are relied on by the whole community; the full scope is necessary for a functioning ecosystem. However, across the different projects and collaborations that compose this field, it is challenging to maintain and update this critical software.  In addition, it is essential to train new scientists in the use of current tools, and to encourage innovation -- there must be space for {\it new} software and tools that benefit the multi-messenger ecosystem.

{\it We recommend a three-tiered system of support for new, emerging and established software infrastructure programs}.   All new software should be open-source, and the expectation should be that it is ultimately published in the literature so that others can build off supported efforts. Such a program will serve multiple purposes, training the next generation of software-savvy scientists while simultaneously sustaining and reinvigorating essential infrastructure:

\begin{itemize}
    \item \textit{Phase 1 (Alpha release for an individual/team):} There must be a mechanism through which new tools -- not conceived of here -- can be developed, tested, and presented to the multi-messenger community.  Often these projects are led by early career scientists with an idea.  We recommend a continuously open solicitation for early career (graduate students and postdocs) researchers to fund 6-12 months of effort ($\sim100$\,K) on MMA/TDA-specific cyberinfrastructure. If possible, this `Alpha' program should include cross-training in software engineering practices that are heavily used in the field. As many viable ideas as possible should be funded in order to ensure a diverse ecosystem of software tools.
    \item \textit{Phase 2 (Beta release for the broader community):} The next step in the development cycle is to deploy successful tools beyond an individual research group/team and to the broader community. For projects that meet their Phase 1 milestones, a cadenced (annual/bi-annual) call for Phase 2 proposals would provide sufficient resources for professional software development and maintenance ($\sim1$M). Sustained support (3--5 years) is critical to the success of such efforts. 
    \item \textit{Phase 3 (Production release for critical infrastructure):} A select few software tools are so critical to the community that they cannot be allowed to be subject to the whims of review for operations funding (e.g., GCN, LSST-scale alert brokers, science platforms). 
    Additionally, for established Phase 2 projects, funding agencies should define a process whereby cyberinfrastructure teams can qualify for and compete to be redirected to a national lab, NASA center, or equivalent (e.g., NOIRLab), the rare organizations that can provide stability on a decade-plus time scale.  
    \end{itemize}

In the following subsections, we outline further recommendations and discuss the full scope of software infrastructure available today.  Whenever possible we point out missing pieces or further infrastructure that is necessary for multi-messenger astrophysics in the decade ahead.

\section{EM Messengers}
\label{sec:SW_EM}
\noindent
{\bf Lead: Michael Coughlin\\
Contributors: Tomas Ahumada, Sarah Antier, Dave Coulter, Brian Humensky, Robert Nikutta, Monika Soraisam, Rachel Street, Bryan Miller, Samuel Wyatt
}\\

\noindent Here we lead with software and archive recommendations related to EM messengers, and then move on to an overview of the current EM software landscape.
\smallskip\\
\noindent \textbf{Recommendations.} 

\begin{itemize}
\item \emph{Advocate for diverse toolkits and computing resources}: different tools and services are essential in different areas of astrophysics, and supporting these diverse sets of tools is required to maximize science returns. The goal for these toolkits is to improve access to information, lower barriers to entry for all within the field, reduce the workload required of observers and operations teams, and create efficiencies across the system. There is also value in centralized toolkits which provide comprehensive capabilities, such as the Gamma-ray Data Tools (GDT), which is setting the foundation for access and analysis of hard X-ray and gamma-ray data.

\item \emph{Encourage interoperability | services and archives}: services are most effective when they integrate seamlessly as a software ecosystem, rather than operate independently. It is inevitable that different services must couple together to achieve many science goals, from the data reduction tools that synthesize the data sets, to the alert brokers that classify and identify discoveries of interest, to the telescopes used to perform follow-up observations, to the messaging services used to disseminate findings. \texttt{Well-documented (and even standardized) Application Programming Interfaces (APIs)} are required for efficient communication between these systems. Agreeing on a common standard single-sign for secure authentication and authorization would be highly beneficial. In addition to ensuring interoperability between cross-agency archives, a \texttt{dedicated MMA/TDA archive infrastructure} that can bring together data from space- and ground-based observatories should be designed.


\item \emph{Teach best software practice}: There are a number of important factors in creating successful software platforms {\it and} collaborations: (i) building open-source and unit-tested software deployable within modern frameworks, such as conda, Docker, and Kubernetes. As NASA migrates to a fully open source paradigm through the Open-Source Science Initiative (OSSI) \footnote{\url{https://science.nasa.gov/researchers/open-science/}}, we suggest code and data products progress accordingly. (ii) Cross-training researchers in software engineering practices and frameworks that are heavily used in the field (e.g., source control, object-oriented design, web frameworks, database languages, and software topologies); and (iii) building in maintenance and management plans to ensure the longevity of software that has become essential.

\item \emph{Pathways to long-term support}: while timelines for initial development of most key services vary widely, it is crucial to provide sustained support for long-term maintenance to, e.g., fix bugs, add new community-developed and tested functions, support for new instruments, and improve documentation, tutorials, and regression testing. One path forward could be the three-tiered system discussed at the beginning of Section~4, where there is a path for critical infrastructure to receive longer-term support, possibly through national facilities with the longevity necessary for such tasks. 
One initiative to highlight here is NASA's Astrophysics Cross-Observatory Science Support (ACROSS), which seeks to partner with observatories and observers to identify capability gaps, highlight best practices, and develop processes and tools that streamline, standardize, or automate coordinated science planning and execution workflows. 
\end{itemize}

\textbf{Community Coordination Services}.
High-profile discoveries can trigger intensive observational efforts by many teams, potentially leading to an inefficient use of resources. This is especially likely given the mismatch between the localizations provided by multi-messenger instruments and the typical follow-up instrument field of view. A number of platforms have been developed to share information and enable teams to coordinate their efforts in real-time.

\begin{itemize}

\item \emph{Heralds }. The General Coordinates Network \citep{SiRa2023} (GCN), with a 30+ year heritage, remains a popular platform by which astronomers both ingest and distribute alerts and follow-up within the time-domain community; it recently has evolved to be built on modern, open-source, reliable, and secure alert distribution technologies. Besides GCN, the Scalable Cyberinfrastructure to support Multi-Messenger Astrophysics (SCIMMA) has built multiple data sharing toolkits, including {\em Hermes}, which provides a robust, low-latency communication service for researchers to share a variety of data types in a machine-readable form that integrates into the wider ecosystem. GCN and {\em Hermes} offer flexible, browser-based interfaces and APIs that enable researchers to receive, compose, and share data via Apache Kafka-based cloud-deployed messaging systems.

\item \emph{Science Platforms}. Astronomical Science Platforms (SP) seamlessly combine big data and co-located computing, alleviating a wide community of researchers from concerns about data transfer, compute capabilities, and intricate software installations. For example, machine learning has become indispensable for MMA/TDA studies, be it classification of real sources from artifacts in the difference images generated by surveys, or vetting of the candidate astrophysical transients to identify the electromagnetic counterparts of the MMA/TDA event. Training of sophisticated deep-learning based classifiers is a compute-intensive task. GPUs are now a requirement to accelerate their training. SPs equipped with GPUs will provide the most efficient route for MMA/TDA researchers to access this compute resource {\it directly at the location of the training data}. Such an infrastructure will particularly be invaluable to researchers in smaller institutions. Funding agencies should therefore support efforts to incorporate GPUs within the computing arsenal offered by SPs. 

SPs can further act as technological glue in MMA/TDA, integrating tools across the entire data cycle from alert stream filtering to triggering of follow-up observations, data reduction, analysis, and publication preparation. NOIRLab's Astro Data Lab \citep{datalab2014,datalab2020} (DL) exemplifies this by hosting the ANTARES \citep{Matheson_2021} alert broker client and filter dev-kit for development of efficient custom filter functions. DL also supports Gemini DRAGONS \citep{dragons-one,dragons-quick} and IRAF data reduction kernels. Future plans include integration with TOMs (e.g. the Gemini Observation and Analysis of Targets System). 
As another example, for the upcoming Rubin Observatory and its Legacy Survey of Space and Time (LSST; \citep{Ivezic2019}), the Rubin Science Platform will co-host its real-time and annual data releases and maintain a software environment that includes the LSST Science Pipelines (\citep{LSE-319,OMullane2021}).

\item \emph{Transient Name Server}. TNS is the official IAU mechanism for reporting new astronomical transients and assigning named designations to new objects.

\item \emph{Coordinating telescope networks}. The Gravitational Wave Treasure Map \citep{Wyatt2020GWTM} is an open-source system for reporting, coordinating, visualizing, and assessing searches for, and subsequent follow-up of electromagnetic counterparts to gravitational-wave events. The ultimate goal is to have every team communicate in real-time through the provided API to optimize their search efforts through passive coordination. Additionally, some telescope networks have adopted a coordination strategy where observations are pre-allocated to optimize coverage at a network level. Networks such as the Global Rapid Advanced Network Devoted to Multi-messenger Addicts (GRANDMA) or the Global Relay of Observatories Watching Transients Happen (GROWTH) have adopted such toolkits to structure their follow-up efforts \citep{CoAn2019}. Both concepts aim to increase the overall efficiency of GW searches by informing, publicly or internally, where and when observations are performed. To mitigate discrepancies in the photometric data analysis, some networks are prototyping platforms by which front-end interfaces to common backend reduction tools are made available to observers  \citep{2021ascl.soft12006K}.

\end{itemize}

\textbf{Brokers and Marshals}.
The dissemination, filtering, follow-up, and analysis of survey data sets rely on two important, interrelated pieces of infrastructure developed within the community: alert brokers and Marshals. Brokers ingest alerts from optical surveys and provide value-added annotations such as cross-matching with other catalogs. Generally, brokers collect and store alert data, enrich them with information from other surveys and catalogs or user-defined added values such as machine-learning classification scores, and redistribute the most promising events for further analyses and follow-up observations. Examples include Fink \citep{MoPe2020}, ALerCE \citep{FoCa2021}, ANTARES \citep{Matheson_2021} or AMPEL~\citep{Nordin:2019kxt}. Target and Observation Managers (TOMs, also known as Marshal) systems, which are downstream of brokers, have proven to be powerful tools for gathering, visualizing, and analyzing information on targets of interest, and managing observing programs. They provide researchers with a single platform that offers interactive and programmable interfaces to key services. Valuable across 
 all domains of astrophysics, they are especially vital for MMA/TDA science where a rapid response is often critical to success.  Examples of TOMs/Marshals developed in the community include YSE-PZ \citep{Coulter_2023}, SkyPortal \citep{Coughlin_2023}, the SAGUARO TOM \citep{hosseinzadeh2023saguaro} and the upcoming Gemini Observation and Analysis of Targets System (GOATS); the last two of these were built with the open source TOM Toolkit \citep{StBo2018}.

\section{Non-EM Messengers}
\label{sec:SW_nonEM}
\subsection{Gravitational Waves}
\noindent
{\bf Lead: Deep Chatterjee\\
}

The LIGO-Virgo-KAGRA fourth observing run is discovering compact binary coalescence (CBCs) on average every $\sim 3$ calendar days at the time of writing. The median latency of the alerts is $\sim 30$ seconds from merger time. In addition, low-significance alerts are also distributed with false alarm rates of 2 per day. The challenge in the online alert campaign is the seamless coordination of several components to discover potential multi-messenger candidates and deliver useful data products. The process includes data acquisition and calibration, data transfer, searching for candidates, archiving candidate information, rapid inference, and annotation before alert distribution through alert brokers SCiMMA and GCN. The coordination of every component of the low-latency system is essential for discovering and delivering GW alerts in real-time. 

In addition, early-warning searches for low-mass systems like binary neutron stars have been deployed during the fourth observing run. Also, new flavors of searches like sub-solar CBC searches, Burst binary black hole (BBH), or searches based on machine-learning and artificial intelligence continue to be added. The distributed nature of the searches, dynamically selecting the best candidates and reporting useful data products for follow-up is a highlight of low-latency GW science.

Due to the cross-talk between the several subsystems, software tools geared toward real-time orchestration are needed to achieve success. Given the ``burst"-like nature of the discoveries and the scope of target of opportunity on the follow-up side, this needs resources that are highly-available. This is not just limited to the availability for compute, but also highly available databases with authenticated and public views, visualization tools, and robust APIs to rapidly perform transactions to keep the \emph{state information} of the physical event to send timely alerts. The low-latency alert infrastructure in the LVK depends critically on three pillars 1) GraceDb: a database that stores candidate information and provides an API to fetch and update the state information of physical events 2) IGWN alerts: A messaging service that uses the SCiMMA Kafka broker under the hood and 3) GWCelery, a task queue platform that contains the business logic to orchestrate the state of physical events, launch inferences and annotation needed before alert delivery.

Due to the broad use cases of real-time applications, several open-source software tools already exist. In the case of the LVK, the orchestration framework, for example, is built on top of open-source real-time software package \texttt{Celery}. The fault tolerant feature, fine-tune worker configuration, and the ease of building workflows are some of the features that were found useful in this case. However, due to the rapidly changing technology landscape, doing periodic technical audits and researching new tools have to be budgeted for in future operations. Easier integration capabilities should also be considered. For example, due to the adherence of tools like \texttt{Kafka} and \texttt{Avro} by the broader astronomical community, these tools provided by the Apache Software Foundation might provide an advantage in this regard. It is also important to budget for end-to-end continuous testing, possibly via mock data challenges -- both from an operational standpoint and the bench-marking of science data products. 
\subsection{Neutrinos and High Energy Particles}
\noindent
{\bf Lead: Hugo Ayala\\
}

As mentioned in Section~\ref{sec:HW_nonEM_neutrino}, high-energy neutrinos can provide a complementary view of the most extreme environments and phenomena in the universe, such as blazars and explosive transients. Since 2010, high energy neutrino alerts from IceCube have been made public. From these, there have been several claimed multi-messenger detections, such as the observational campaign associated with the blazar TXS0506+056, which was also observed in gamma rays by the Fermi Telescope \citep{nugammaTXS}. Another alert was plausibly associated with a tidal disruption event \citep{tdeIceCube}. However, in the case of these IceCube alerts, there can be a delay of the order of hours before the best reconstruction algorithms can be applied to the alert events and are sent to the public. The second revision of the alert requires human intervention and vetting before being sent to the public. Several enhancements to the real-time alert system for neutrinos are available to improve these alerts. First, improved real-time reconstructions that can provide higher accuracy event direction and energies promptly can be developed that operate at the Antarctic detector site and removing the need of a second alert revision. This, coupled with the need for additional computer resources at the South Pole, will enable rapid and accurate reconstruction and classification for a larger number of neutrino events for use in MMA/TDA analysies. Second, additional software is needed to  automate the rapid integration of these neutrino alert events into community alert and broker systems. These software systems would automate the alert communication using community standards (such as FITS files for event localizations as done by the LVK group), and rapidly distribute them to allow rapid response or reaction by other MMA/TDA observatories. 

\section{Synthesis of Multi-Messengers}
\label{sec:synthesis}
\noindent
{\bf Lead: Gautham Narayan\\
}

As detailed in previous sections, much of the software infrastructure that is necessary for EM, gravitational wave, neutrino and high-energy particle observatories to communicate their discoveries already exists. Users can already subscribe to LVK alerts via e.g. SCiMMA, process data embedded in the skymap using e.g. NOIRLab's Astro Data Lab, and trigger spectroscopic follow-up with AEON.  Further advertisement of these capabilities are necessary, as are published examples showing these somewhat disparate tools used in a coherent way. However, the rate of events yielding true multi-messenger astrophysical discoveries (i.e. with detections from multiple facilities and experiments using different probes) is low  \citep[][]{CounterpartRate23}. 
The detection of any of these rare events by a single probe can be viewed as starting a stop clock, beginning an all-to-brief window in which the community needs to collect as much information as possible before these rapidly evolving events fade below detectability. Consequently, it is necessary but not sufficient to have each facility  share its data with the community - we will need to \emph{combine and synthesize} information in real-time to coordinate community efforts and maximize scientific return. 

Given the diversity of new facilities coming online in the next decade (Rubin, Roman, A+, etc.), we consider each step in the process of synthesizing MMA/TDA observations below, and make these recommendations, highlighting if they are associated with the funding agencies, facilities, the community, or a combination thereof:
\begin{enumerate}
    \item \textbf{Funding agencies:} New facilities will be reluctant to adopt any tools or standards that do not have a stable funding stream for the duration of the experiment, often preferring to develop an internal version that they are confident can be maintained. This leads to a fractured ecosystem. Consequently, we recommend that critical community-wide infrastructure be moved to Phase 3 ({\it Production release}) to facilitate this.
    \item \textbf{Funding agencies:} While ensuring long-term support for tools and infrastructure can help their adoption, we also recommend that NSF and NASA use their often overlooked data management plans to ensure that proposed experiments adhere to incorporating these tools, and maintain a central resource of relevant funded projects that proposal teams can refer to during the  development process.
    \item \textbf{Facilities:} Synthesizing the data from different probes such as gravitational wave interferometers and ground-based imaging is intrinsically a challenge of processing \emph{multi-modal data} in real-time. Combining imaging data from multiple ground-based telescopes with different fields-of-view, filter sets, pixel scales, and site properties is a complex, unsolved problem by itself, let alone combining data across different multi-messenger probes. Tackling this challenge requires reasonably mature data reduction pipelines to process ground- and space-based data in real-time, and disseminating these reduced data products to the community. 
    Funding agencies should require observatories to provide pipelines for processing data \emph{automatically} from instrument commissioning. 
    \item \textbf{Facilities:} Observatories and other research facilities must modernize and democratize data access from their existing archives, e.g. by ensuring researchers can access datasets via API, and automatically building public collections of datasets for each MMA/TDA event both from archival observations, and from any follow-up observations. As mentioned in previous sections, we recommend dedicated MMA/TDA archive infrastructure that can facilitate this.   
    \item \textbf{Funding Agencies and Community:} The rarity of multi-messenger events, and the multi-modal nature of the data are challenges that require different and new astrostatistical techniques. For example, uncertainty quantification is significantly more challenging across multiple facilities, each with their own sensitivity and processing pipelines, and the inherently small number statistics makes e.g. an analysis of rates challenging. 
    AI will play a major role in the next decade in coalescing these multi-modal datasets, and retrospective analyses of MMA/TDA campaigns will likely provide an evolving picture of e.g. the neutron star population, and their connection to their host-environments. These efforts will need funding, but will also require data \emph{and metadata} to be preserved and documented rigorously for periods longer than a typical graduate student thesis.
    \item \textbf{All:} Synthesis of MMA/TDA observations should be viewed as an ongoing conversation. We recommend a standing joint expert group and public telecon, with liaisons for each major MMA/TDA experiment to foster communication between these presently-siloed experiments. This could naturally evolve from the existing OpenLVEM series of telecons. We anticipate that a shared forum with stakeholders from the facilities who are receptive to feedback from the community will ultimately foster more collaboration and make the MMA/TDA ecosystem richer.   

    \end{enumerate} 

\chapter{MMA/TDA Infrastructure Challenges and Proposed Solutions: People and Policy}\label{sec:PP}

\noindent
{\bf Lead: Rachel Street\\
Contributors (alphabetical order): Tomas Ahumada, Sarah Antier, Adam Brazier, Poonam Chandra, David Coulter, Adam Goldstein, Leanne Guy, Eric Hooper, Shaniya Jarrett, Jamie A. Kennea, Tiffany Lewis, Lu Lu, Tom Matheson, Bryan Miller, Gautham Narayan, Krystal Ruiz-Rocha, Monika Soraisam
}

\section{Inclusive Workforce Development}

Fostering a diverse and equitable workforce is essential to promote a vibrant science community and to strengthen the foundational infrastructure needed for MMA/TDA research. 
Ensuring that people of all ethnicities, nationalities, gender identities, orientations, ages, disabilities, economic backgrounds, and career stages feel valued and supported ensures their maximum productivity and the retention of expertize within the field, as well as fostering innovation through cross-disciplinary communication.  
While the challenges, such as systematic bias, underrepresentation, limited resource access and more, are not unique to multi-messenger astronomy, they are particularly impactful as it is an emerging field.  

Some institutions can offer their research staff considerably more resources than others, including funding,  computing and IT infrastructure.  This creates an uneven playing field in terms of research output and staff time commitments.  Unequal teaching loads were highlighted as a particular constraint, particularly at smaller institutions, and those serving communities historically under-represented in astrophysics.  The inequity of opportunity extends to students as well, since schools with more resources can send their students to conferences or on talk tours.  

NSF initiatives like the CloudBank \citep{cloudbank}, and the Rubin Science Platform (RSP) \citep{LSE-319, leanne_p_guy_2021_5683849}\footnote{\url{http://data.lsst.cloud}} which provide community access to Cloud computing resources, are a valuable mechanism for democratizing access to data and computing services. These should be supplemented by providing funding channels aimed at supporting on-site computing and IT infrastructure for research groups where this is not provided by the institution.  

NASA and the NSF are in a unique position to use their influential Town Hall platforms to encourage institutions to issue calls for abstracts to make their invited talks/colloquia series more open.  Offering a relatively small funding channel to cover speaker travel costs in exchange for these open calls could incentivize institutions to make these opportunities more accessible to a more diverse group of students.  

While some level of competition is recognized to be healthy, hyper-intense competition between different groups in the field was recognized to disproportionally favor aggressive people to the detriment of those who are more reserved, who can be overlooked or sidelined despite their equally valuable research contributions.  This disincentivizes science, deters participation, and rewards uncollegial behavior.   Historically, this has disproportionally affected women, and Black, Indigenous and People of Color, who face greater social pressures deterring aggressive or assertive behavior \citep{Motro2022}.  

These challenges should be mitigated with a number of strategies:
\begin{itemize}
    \item Implement an agreed Code of Conduct for MMA/TDA research, outlining expectations regarding data sharing and co-authorship in a public document. 
    \item Consider developing a standard ``data license" (analogous to public intellectual properties licenses such as the GNU General Public License).  In concept, this would enable data to be shared and used for real-time analysis purposes, but require (e.g.) co-authorship or the appropriate citation for publication.   
    \item The creation of a body to adjudicate impartially in cases of dispute. 
    \item Actively foster and reward partnerships between larger and smaller institutions and those serving underrepresented communities, for example by funding sabbaticals, and student visits.  
    \item Prioritize inclusive training alongside research, encompassing topics such as mentorship for minority students, platforms for underrepresented groups to influence decision-making, workshops on unconscious bias, and gender equity policies.  

\end{itemize}

Concepts for an MMA/TDA Code of Conduct and/or a data license led to extensive discussions.  It would be beneficial to organize a further community meeting to continue these discussions and to arrive at a widely-supported solution.

Open data is especially important to enable research at  under-resourced institutions.  That said, open-data policies need to be thought through carefully to avoid potential negative impacts, for example on junior researchers.  The data policies for survey versus follow-up facilities should be considered separately, given the different science goals and potential for ancillary science with the data products.  If PI-level projects are required to share their data, then funding should be provided to enable the data to be prepared and documented for release, as well as to national-level archival facilities to host the data long-term.

\section{Education and Training}
Fostering programs that enable the exchange of techniques and tools between teams and fields of research, as well as between industry and academia, offers a number of benefits: overcoming `siloed' thinking, cross-fertilizing research by sharing ideas, and providing a conduit for innovations in academia and industry to benefit both domains.  This is particularly important for MMA/TDA, since industry developers have had to deal with  Big Data for considerably longer than astrophysics.  It is therefore recommended that venues be developed to provide broad access to training on tools and techniques that are widely advertised and accessible to the whole community.  These venues should include support for $\sim$week(s)-long training schools, analogous to the Zwicky Transient Facility Summer School, but broadened in scope, as well as funding internships and sabbaticals.  Partnerships between academic institutes and industry would not only train researchers but also provide researchers with a broader perspective on career options.  
Funding should also be provided to enable curriculum content to be developed, and made freely available online.  The Software and Data Carpentries \citep{sw_carpentry} offer an example of a successful model.  

The Establishing Multi-messenger Astronomy Inclusive Training (EMIT) program\footnote{\url{https://www.vanderbilt.edu/emit/}} was recognized as a valuable workforce training program designed to address disparities in student training and experience.  EMIT is a 2-year certificate program that students can complete as a 2-year Master to Ph.D. bridge program or in parallel with their first 2 years of their Ph.D. Students are required to complete 2 core courses, 2 electives, and two “blast modules.” The program can be tailored for different career goals (industry, civil service, academia).

\section{Career Paths}
Effectively realizing the scientific potential of MMA/TDA necessitates developing, maintaining and operating high-quality software infrastructure.  In order to accomplish this, there must be clear and widely-recognized career paths for software engineers in astrophysics.  These must be designed to appeal to and to retain both software engineers who work in research and astrophysicists with an interest in software.  Experience from a number of software infrastructure projects (e.g. the TOM Toolkit, \cite{Street2018}, SCiMMA-SNEWS, \cite{Baxter2022}) has demonstrated the value of partnerships between these groups in terms of delivering better quality, more sustainable code for tools that are tailored to the needs of research.  

There was a widespread community perception that too many institutions do not recognize or reward contributions to infrastructure, for example in terms of awarding tenure.  The development of a vital software package or instrument might result in relatively few traditional journal papers (or technical papers such as SPIE contributions), and yet be indispensible in making a wide array of science possible.  Attitudes and reward mechanisms must change in order for our field to attract and retain highly skilled software engineers.  Funding agencies can most directly influence attitudes by offering funding avenues explicitly designed to support these roles, and recognizing their products.

This funding must provide specific avenues for the long-term maintenance of vital software, as such proposals inevitably lose out to novel, short term science goals in traditional funding streams.  Short-term or soft-money contracts for software engineers puts academia at a severe disadvantage when trying to recruit and retain engineers in competition with higher-paid industry positions.  Funding channels must be provided to enable software engineers to be employed in research groups long term.

To further encourage the community to value and reward software developers, agency representatives should particularly highlight the value of these roles in proposal call language, Town Hall and other presentations to the community.  
Another opportunity to foster change is to offer postdoctoral fellowships that specifically encourage software as well as science focus. The evaluation of applicants would need to reflect a greater weight on development, engineering and operations than the traditional emphasis on Nature papers.  

It is important to recognize that career paths start in grad school, and there is a danger that if support for software developers occurs only from the post-doctoral level and up, academia will lose talented researchers before they have a chance to flourish.  NASA and the NSF can spur a cultural change in institutions by explicitly recognizing a third path in grad school - not just instrumentation vs science (and observation vs theory), but also ``scientific software engineering''.  National standards could be outlined to set expectations on the training and career recognition for graduates in this channel.

\section{Removing redundancies at Observatories}
There is enormous load on major facilities for follow-up of the EM counterpart candidates of multi-messenger events. For example, more than 240~hours were requested on Gemini in semester 24A for very similar science programs (14 proposals in total{\footnote{\label{noir_TAC}We thank the Gemini ITAC chair (Mark Rawlings) and NOIRLab TAC program head (Verne Smith) for providing us the statistics.
}}), namely following up gravitational wave events from the LVK O4 run using the same instruments and modes. Obviously, duplicating observations is not the best use of valuable 8-m time, and could have collateral damage to other fields of astronomy by reducing their share of time. The community {\it in collaboration with observatories} should instead focus on eliminating redundancies and getting the best data possible to extract the maximum science. To initiate this, observatories like Gemini should establish a key project using Director's Discretionary Time (DDT) for follow-up of multi-messenger events where the \underline{data are immediately public}. The Back with a Bang program{\footnote{https://noirlab.edu/science/index.php/news/announcements/sci23033}}, which was a Gemini DD program for observations of the closest supernova in nearly a decade, SN~2023ixf, is an example of this model. The resulting data were immediately used in several publications \citep[e.g.,][]{Pledger2023, SVD2023}. The STScI ULLYSES DDT project for HST is another example\footnote{https://ullyses.stsci.edu}.  

To this end, the observatory should invite MMA/TDA experts from the community to define the scope of the program -- instruments to be used, trigger criteria, cadence, etc. New developments in software infrastructure (discussed in Section~\ref{sec:SW_EM}) could facilitate such an experiment. Ideally, the key project should be open to the entire MMA/TDA community. The program can set up a rotating executive committee for defining and updating its policies, including identifying people who can trigger, credit and authorship in publications, etc. The proposal for the key project could be published as a living document, where new members can be added, and the data from the project be licensed requiring citation of this proposal in any publications using those data.

Observatory staff would certainly play a critical role in bringing such a project to fruition. As such, facilities should pull out all the stops to hire and retain researchers with experience in this field -- ensuring they also remain competitive to be able to understand the needs of the community and thus be able to better support them. On the other hand, funding agencies, in particular NSF, should consider adopting a model where the time allocated for its facilities via the NOIRLab Time Allocation Committee (TAC) is backed up with a small amount of funding to support investigators. These funds could be used by PIs to complete their programs and to publish and share the results at conferences and workshops.   

We recommend implementation of key projects for MMA/TDA follow-up at ground-based observatories operated by NSF in the near term, while funding for TAC-approved observing programs on these facilities can be rolled out in the near- to mid-term. Based on the statistics available for the NOIRLab TAC{\footnote{https://noirlab.edu/science/observing-noirlab/proposals/tac; see also footnote \ref{noir_TAC}}}, 
roughly 200 programs are allocated time each semester. Assuming a grant of a few thousand dollars per program, sufficient to cover at least publishing and conference travel costs, would amount to only a few million dollars per semester.

\section{Overcoming Barriers and Silos}
The workshop highlighted complementary research that is currently occurring in isolation, and where coordination between different groups would benefit the science return and improve efficiency. 

Firstly, both NASA and the NSF have established MMA/TDA programs and scientific interests.  To a certain extent, these programs, and the infrastructure associated with them is necessarily distinct - for example, the operation of space-based observatories relative to ground-based ones.  Nevertheless, similar challenges arise, and where appropriate, NASA and NSF planning, programs and infrastructure should be designed to be compatible and to emphasize interoperability.  This requires close, long-term and regular coordination across agency boundaries but offers a number of benefits to the science, including:
\begin{itemize}
    \item Proposal calls: MMA/TDA science requires access to resources (archives, observing facilities, alert streams, etc) that are operated by both NASA and the NSF.  To write a well-motivated proposal, researchers must request access to data and facilities across different facilities.  The burden on researchers is reduced (and the likelihood of success increased) if they can request all the necessary resources in a single proposal, rather than submitting related piecemeal proposals to separate agencies.  The cross-over program between Chandra and the Hubble Space Telescope is an example of the benefits of this kind of initiative. Coordination at the agency level would help to time these calls in advance of key opportunities, such as the launch of new missions or the start of a new survey.  

    \item Hardware and software infrastructure: The agencies should coordinate plans for development so that there is no lapse in the access of US researchers to instrumentation across the wavelength range (see Sections~\ref{sec:HW_EM_radio}--~\ref{sec:HW_EM_Xray}) to ensure success of MMA/TDA science. 
    As a community we share a vision of the future of MMA/TDA, wherein neutrino detectors, GW detectors, and electromagnetic detectors are all simultaneously observing with sufficient and complementary sensitivity such that when the universe provides information, we have the ability to collect it. It is {\it not} merely a matter of collecting all the resources to do the best science and if we are missing 10\% then we still get 90\% of our goals. Some of our scientific goals are simply not possible without a minimum complete set of instruments and the coordination infrastructure to make them work together without delays. 
    There must therefore be an avenue of solicitations or proposal evaluation criteria that allows for {\it coordination as a distinguishing characteristic}.

    It is also vital to invest efforts into making software services run by NASA or the NSF interoperable (Section~\ref{sec:SW_EM}) and to foster sharing of information on the infrastructure available and under development by both agencies, both between the agencies and the science community. Pioneering work to enable time-domain observations has been made at both NASA-run (e.g. {\em Swift}'s Target-of-Opportunity API) and NSF-run (e.g. the AEON Network) facilities.  Voluntary but unfunded coordination between these programs already exists, leading to compatible architecture design - these efforts should be fostered and supported.  Updating the programs to enable observatories to broadcast a conditional readiness to accept ToO interrupts (beyond yes/no) based on type of messenger and event priority would allow more effective and efficient use of facilities with comparable capabilities. Expanding the application of this API to other space-based facilities (as proposed in NASA's ACROSS program) would greatly enable the rapid and effective follow-up of MMA/TDA discoveries.  

    \item International cooperation: To date, the NSF and NASA have negotiated separately with international agencies (e.g. the European Space Agency, Japan Aerospace Exploration Agency) regarding access to data from key missions such as {\em Euclid} or {\em ULTRASAT}.  Since these resources are of interest to a broad section of the community, these efforts should be coordinated to avoid conflicts of interest and miscommunications. Additionally, US-based initiatives should also be coordinated with, and benefit from, other MMA/TDA-related work going on overseas.  For example, the European Opticon/Radionet Pilot program are developing a common infrastructure to enable programmatic observation requests to be submitted to facilities across many different wavelengths, including optical, NIR and radio facilities.  This infrastructure is already AEON-compatible, and time on many European facilities will be accessible to the US community through the Rubin International In-Kind Program.  We should capitalize on the opportunity to build compatible infrastructure at US observing facilities at all wavelengths.  This will enable time-domain programs frictionless access to a wide array of observing facilities worldwide, but its implementation will depend on long term collaborative efforts between US and international observing facilities. 
    
\end{itemize}

A second aspect to siloing in MMA/TDA occurs (unintentionally) across wavelength and messenger boundaries, due to specializations within those communities.  This has the undesirable consequence of raising barriers to entry for researchers who wish to, e.g. incorporate data from an unfamiliar passband into their analysis.  There can be a steep learning curve associated with the understanding where to find the data products, how to access them, and how to properly use specialist data pipelines and analytical tools. Funding a series of workshops and meetings designed explicitly to share information on data handling, observations and analytical tools and techniques between wavelength and messenger regimes would help to overcome these barriers.  Additionally, funding a committee or institute to determine how the NASA, NSF, and other archives can better work together to serve the community, specifically for MMA/TDA science, as suggested by the 2023 FADI Report~\citep{FADI2023} would be invaluable.

Open-source and open-access platforms are a vital resource enabling researchers from all backgrounds to access large scale survey data products and advanced software tools.  It is important that these platforms be designed in such a way as to enable their use by researchers at smaller institutions.  Merely making code open source is often insufficient to enable it to be widely adopted in practise - detailed documentation and training materials must be developed.  This should be made the expected standard for federally funded projects, and developers should be funded to provide these materials.

The Astrophysics Source Code Library \citep{ASCL} and a number of other scientific code repositories maintain scientific and technical codes, including searchability, uniform standards for documentation, and easily accessible citation information for the code itself, not just an associated paper. There should be a way to index citeable codes and keep track of associated citation counts through services like NASA/ADS, arXiv, and InspireHEP. Wide adoption through these services and the expectation from prominent journals that people cite code is one way to address the discrepancy in work accounting between different specialities within the field. Funding agencies could support the necessary updates to current infrastructure as outlined above, but they could also set the expectation that code developed, which they often already require to be published in some form, also be made citeable. In principle, even codes which are not fully public could have public citation information available.

\section{Outreach, Public Engagement and Advocacy}

The general public's interest in and excitement for science are the backbone of support for the continuation of progress in science, through societal funding.  Public engagement also serves a key role in recruiting future scientists, and to raise scientific literacy in the wider population. MMA/TDA, with its groundbreaking and rapidly evolving discoveries, offers a fertile ground for outreach and education. It is a responsibility of publically-funded research to effectively communicate MMA/TDA and its various points of discovery to society at large. 

There is, however, a dearth of ready-made, general-public-level outreach curricula (talking points, hands-on activities, quality graphics etc.) for high-energy astrophysics and MMA/TDA. It should be a goal of broader impacts and DEIA initiatives to make MMA/TDA more accessible through the strategic development (including documentation, workshop, etc.) and dissemination of new materials and resources (via training of other scientists and science students to use these). While there is a cultural resistance among some scientists to spending time on non-science activities, funding agencies have the power to dictate what is valuable within the field by what they offer money and other resources for scientists to do.

Citizen Scientists have a long and impressive track record of contributing astronomical observations, especially through organizations such as the American Association of Variable Star Observers (AAVSO)\footnote{https://www.aavso.org/}, and the Kilonova-catcher program\footnote{http://kilonovacatcher.in2p3.fr/}.  As good quality CCDs and spectrographs come down in price they have become more accessible to the wider public, and there are now hundreds of amateur observers worldwide who are keen to contribute to research programs.  Platforms should be developed to build partnerships between researcher teams and the Citizen Science community to enable them to conduct observations and to share and process their data (beyond the current scope of e.g. Zooniverse \cite{Zooniverse}). This might also incorporate special calls for training.  This is a valuable vector for inclusion in science that particularly appeals to rural communities (who benefit from darker skies than in cities), and yet who often feel neglected by academia.  

MMA/TDA data products, especially alert data, could form the basis of many data exploration and visualization tools designed for the general public, and leveraging innovative tools such as augmented reality.   These resources can be based on existing platforms but still take significant time and effort to produce, so specific funding avenues should be provided for this.  Strong emphasis should be placed on building general-purpose platforms and tools, with good support and documentation, to enable them to be re-used by different community groups without reinventing the wheel.   Facilitating partnerships between domain specialists in the community and agency-level programs e.g. NASA's Data Visualization Studio, would help to develop high-quality content, as well as a widely-recognized avenue for reaching a broad audience.  

Care must be taken to ensure that MMA/TDA outreach programs engage with different social demographics equitably, and that access to the tools and platforms designed for MMA/TDA public engagement do not have hidden barriers.  For example, data visualization tools that require specific hardware may prove too expensive for many people whereas cellphone applications are more broadly accessible.  The unequal access to high-speed internet services should also be considered in the design of these tools.  

Scientists engaging with the media and the public is highly beneficial but few scientists receive any training in communicating effectively.  In an era of political polarization, and where some groups in society are more likely to receive aggressive responses on social media than others \citep{Pew2021}, it is all the more important that researchers have access to media training.

\section{Recommendations}
In light of the considerations above, we make the following recommendations:
\begin{itemize}
\item Bring together stakeholders from the community, publishers and agency representatives to discuss a Code of Conduct and/or data license to facilitate ethical collaboration and healthy competition within the field while ensuring appropriate credit is given, and to address concerns regarding open data policies.

\item Closer coordination between NASA and the NSF is strongly encouraged, particularly in long-term planning (e.g. for facility/mission/capability development) and in negotiations with international partners.  The agencies should support existing efforts to enable MMA/TDA science at NASA and NSF observing facilities through the ACROSS program and AEON Network and fund the extension of these initiatives to radio and other wavelength facilities.  This should be coordinated with parallel and compatible efforts internationally, such as Europe's Opticon/RadioNet Pilot.

\item Invite the MMA/TDA community and observatory operators to design observing programs for MMA/TDA targets where the data would be immediately public.  This should be coordinated with international facilities.  

\item Fund a variety of avenues for training to enable researchers to plan, obtain and analyse data from different wavelengths and messengers.  These should include training schools, internships and sabbaticals, and the development of online training materials.  Programs such as EMIT are vital to foster a diverse and inclusive workforce, and to open multiple career paths to researchers. 

Funding channels must be provided to enable software engineers to be employed in research groups long term (5+yrs minimum).  Rewarding career paths must be enabled to retain software engineers in research as well as researchers with an interest in software development. Signal greater respect for researchers with this specialization in all agency presentations to the community and by weighting software contributions as well as publications in fellowship selection panels. This will ensure the development and longevity of MMA/TDA infrastructure.

\item Funding avenues should be provided to enable researchers to develop public outreach content and tools for MMA science, as well as providing access to media training.  Partnerships between researchers and agency programs such as NASA's Data Visualization Studio should be fostered.  
\end{itemize}

\chapter{Conclusions}
\label{sec:conclusions}
In this document, we have presented recommendations driven by feedback from the MMA/TDA community on how to establish the infrastructure for a collaborative multi-messenger ecosystem. As reflected by the structure of the document, the recommendations can be grouped into three primary themes: hardware, software, and people \& policy.  It is important to emphasize how the MMA/TDA community is faced with challenges exceeding those encountered in other fields, requiring coordination across multiple cosmic messengers and communication across a community with diverse technical and cultural backgrounds. The engagement and feedback received in the October 2023 workshop and the collective effort that has led to this report demonstrate that the MMA/TDA community is determined to address these challenges and ensure the MMA/TDA field reaches its full potential over the next decade. We encourage continued discussion in future workshops that can also expand on the scope of this document, addressing international collaborations and topics such as theoretical work, simulations, and experiments to help interpret MMA/TDA observations.

\newpage
\section*{Acknowledgements}
We would like to acknowledge the workshop participants, SOC, LOC, and all of the people who have supported the workshop and the writing of this white paper. 
The workshop would not have been possible without the logistical, organizational, and technical support from Jessica Harris, Brittany McClinton, Yuanyuan Zhang, David Jones, and Lamont Payne.
We also acknowledge NOIRLab's Communications, Education, and Engagement team for their support of the workshop as well as Peter Winters for his coordination efforts with the workshop venue. Lastly, we thank the staff at the Westward Look Wyndham Grand Resort and Spa for their hospitality and providing an excellent venue for the workshop.  
This material is based upon work supported by the National Science Foundation under Grant No. NSF AST-2336094.

\pagebreak
\appendix

\chapter{Workshop Information}
{\bf Windows on the Universe: Establishing the Infrastructure for a Collaborative Multi-messenger Ecosystem}

The advent of gravitational-wave and particle detectors, which now routinely observe events through new windows upon our dynamic Universe, has ushered in the era of Multi-messenger Astronomy (MMA). With a diverse and powerful network of ground- and space-based instruments and facilities, we now have advanced resources to both identify the electromagnetic counterparts of Multi-messenger events, and to monitor and characterize their evolution. This activity requires coordination of the full range of available telescopes and their capabilities. While the scientific potential is staggering, future campaigns will be resource-intensive, expensive, and require considerable coordination, collaboration, and communication among the communities in order to deliver effective science.

To foster the infrastructure for a vibrant and collaborative MMA ecosystem, NSF's NOIRLab, in partnership with NSF and NASA, invites the international community to participate in a workshop in Tucson, AZ during October 16–18, 2023. The workshop goals are to identify pathways that increase the coordination of observational MMA campaigns and reduce operational redundancy across the network of ground- and space-based observatories. We invite the community to review the current state of resources for MMA, report on existing collaborations and partnerships, and identify potential obstacles to success. We ask the workshop participants to conceptualize community strategies and methods of tackling data sharing, such as telescope pointings, both planned and executed, observation outcomes, and calibrated data. Furthermore, we ask the workshop participants to identify pathways which will incentivize coordination, collaboration, partnership, and collegiality within the MMA community.

Key Questions to Address:
\begin{itemize}
\item{What are the main challenges to perform successful MMA campaigns and to maximize their scientific potential?}

\item{How should we coordinate MMA follow-up to reduce operational redundancy across the network of ground and space-based observatories?
How should we foster collaboration in the MMA community?}

\item{How can we ensure that the MMA field reaches its full potential over the next decade?}

\end{itemize}

Workshop participants will lead the preparation of a community-driven white paper (this document) that will be used to guide NSF, NASA, and NOIRLab planning around fostering the infrastructure for a collaborative MMA ecosystem.

\begin{figure}
    \centering
    \includegraphics[width=\linewidth]{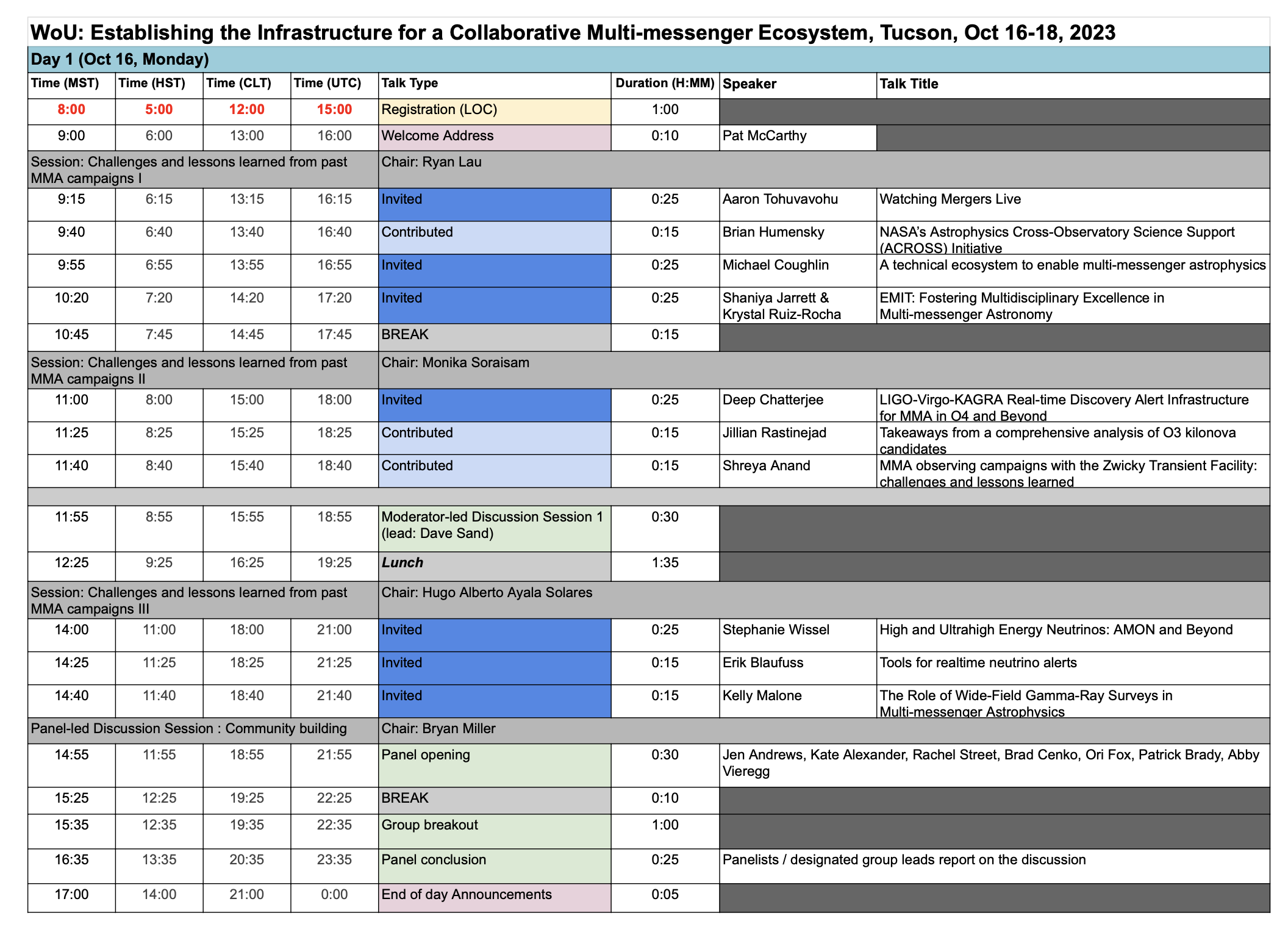}
    {\caption{Day 1 Agenda}}
\end{figure}

\begin{figure}
    \centering
    \includegraphics[width=\linewidth]{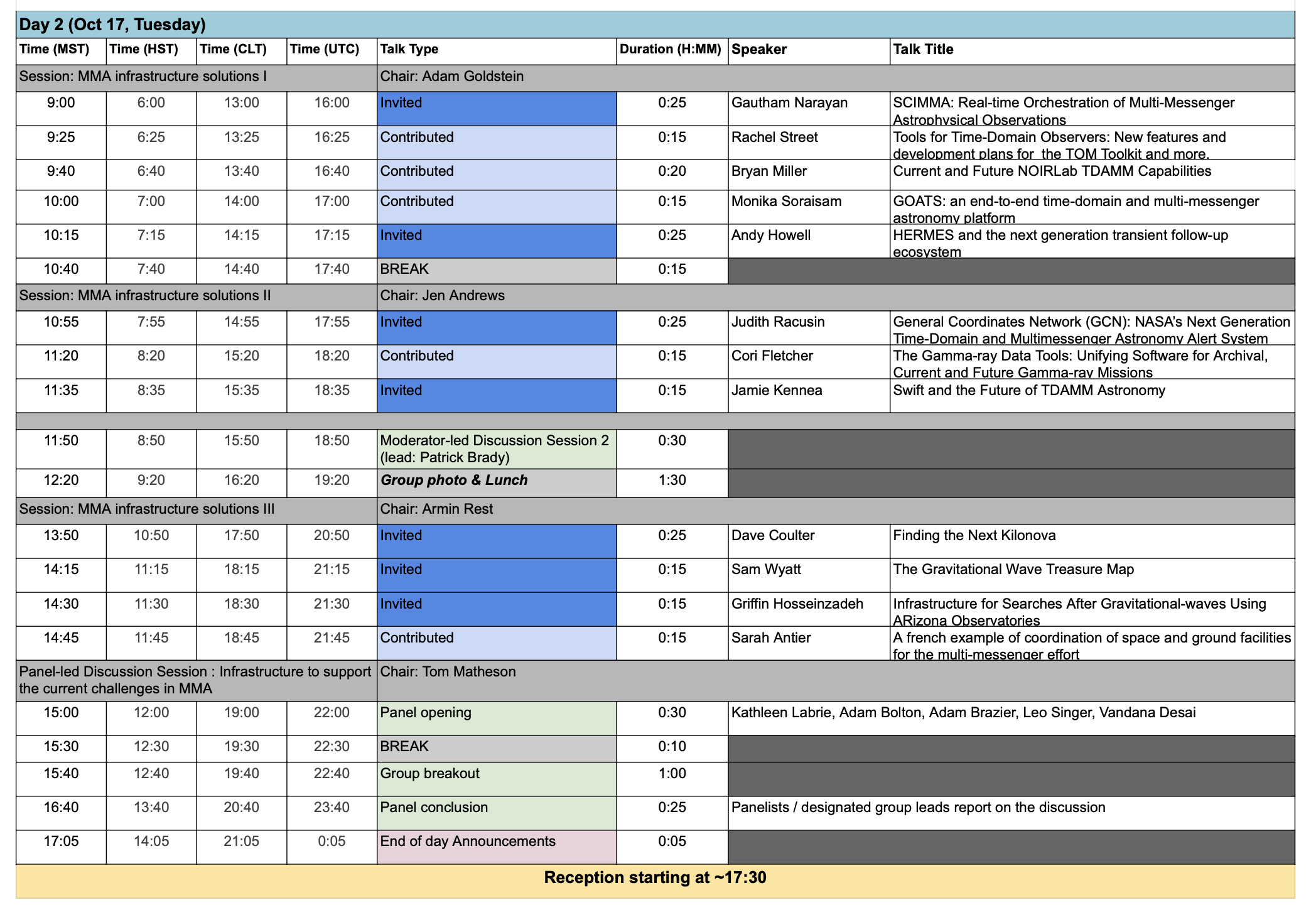}
    {\caption{Day 2 Agenda}}
\end{figure}

\begin{figure}
    \centering
    \includegraphics[width=\linewidth]{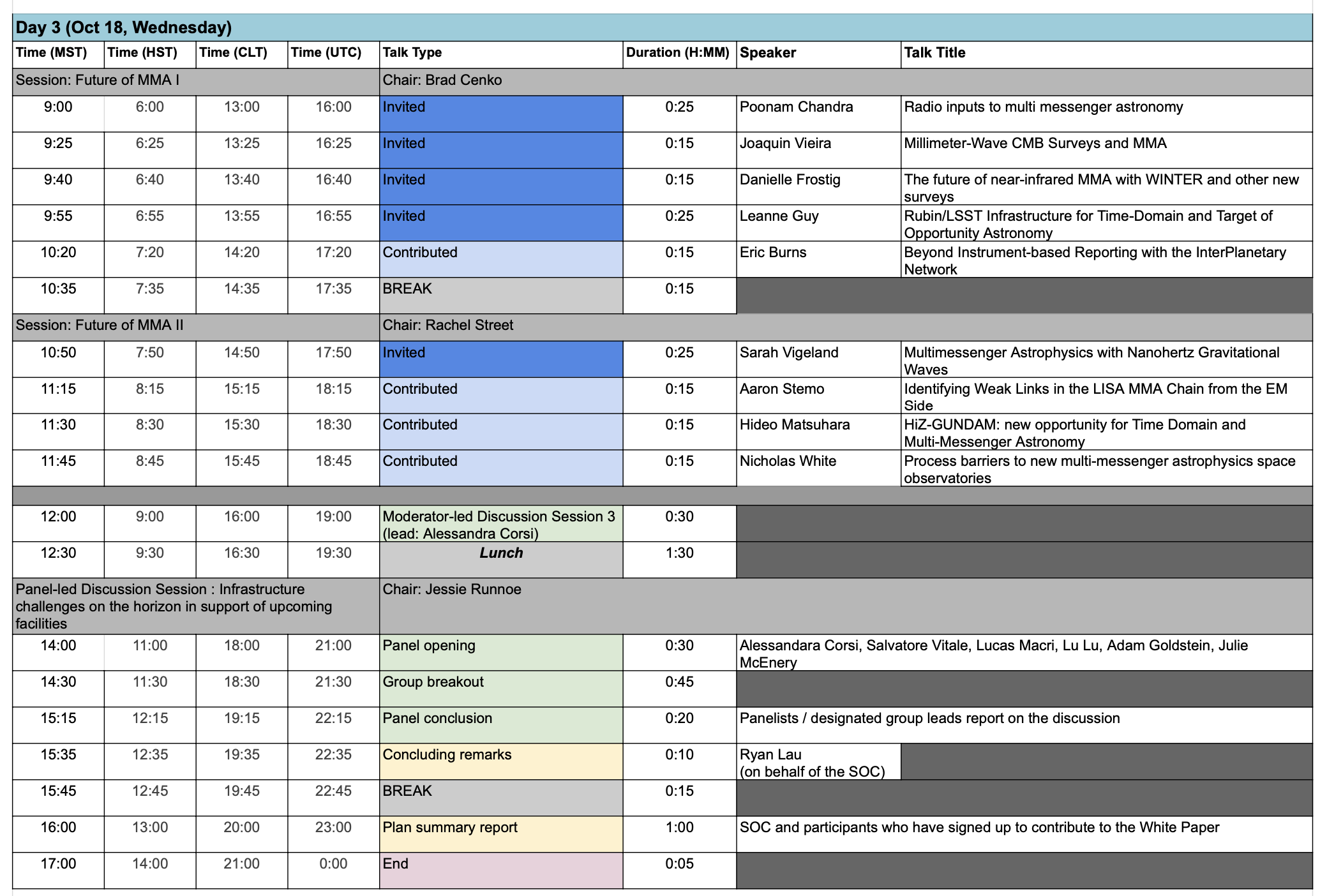}
    {\caption{Day 3 Agenda}}
\end{figure}

\clearpage
\section{Attendance and Endorsement}
\begin{figure}[h!]
    \centering
    \includegraphics[scale=0.71]{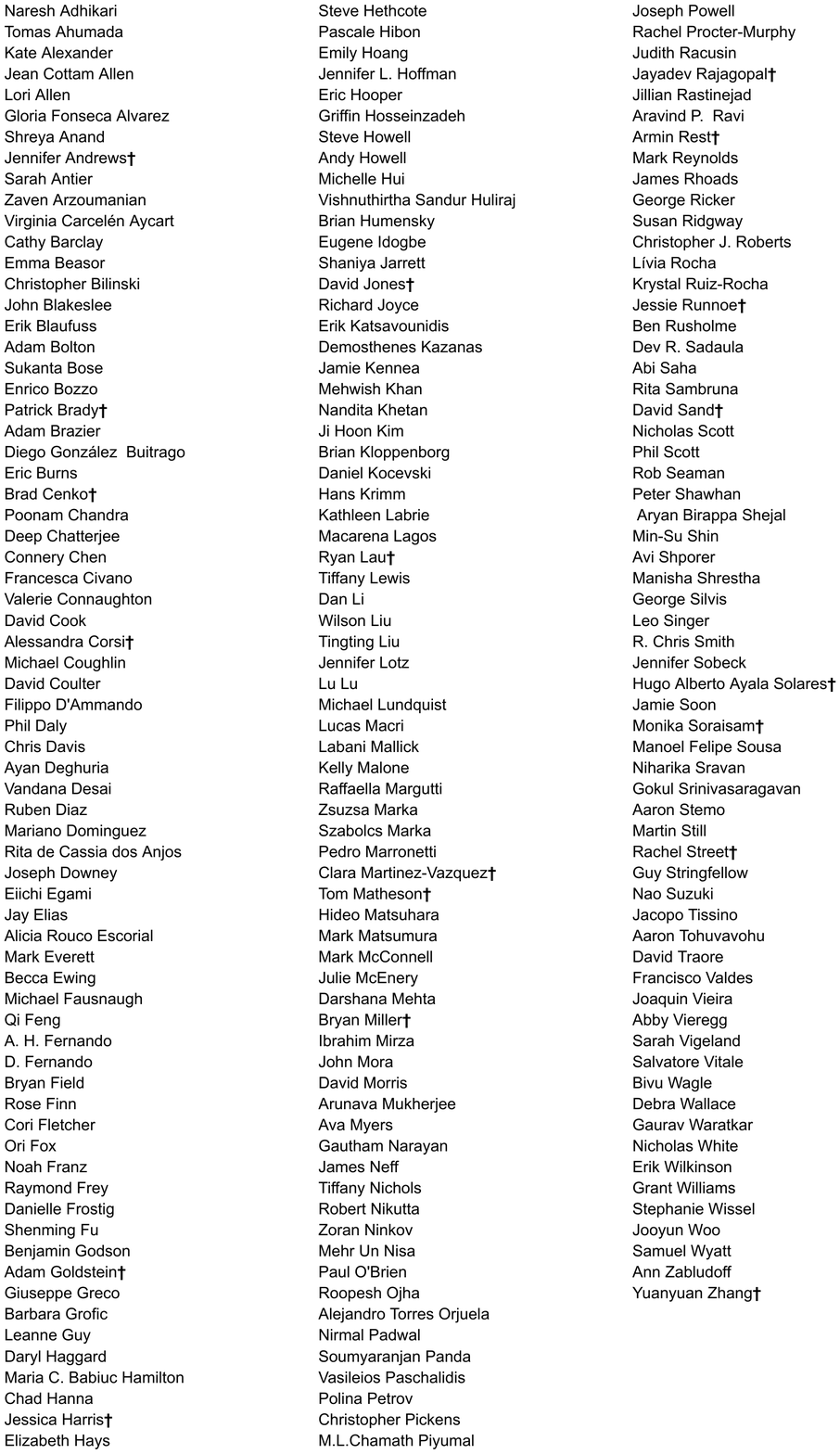}
    {\caption{List of workshop attendees. Dagger indicates SOC/LOC member. List of endorsements to be kept live at \url{https://noirlab.edu/science/events/websites/MMA2023/WP-endorsements}}}
\end{figure}
\clearpage

\bibliography{references}{}

\end{document}